\pdfoutput=1
\newif\ifjournal
\newif\ifwias
\newif\ifarxiv
\newif\ifpdf
\journalfalse
\wiasfalse
\arxivtrue
\pdftrue

\ifjournal
    \documentclass[aip,apl,reprint,a4paper,superscriptaddress,floatfix,amsmath,amssymb,amsfonts,noshowpacs,longbibliography]{revtex4-2}

    \usepackage[T1]{fontenc}
    \usepackage{lmodern}
    \usepackage{amsmath,amsfonts,amsthm}
    \usepackage[utf8]{inputenx}
    \input{ix-utf8enc.dfu}
    \usepackage{graphicx}
    \usepackage[dvipsnames]{xcolor}
    \usepackage{xspace}
    \usepackage[textwidth=17.5cm,textheight=23.7cm,marginparwidth=2cm,verbose,pdftex]{geometry}
    \usepackage[pdftex]{hyperref}

    \hypersetup{pdftitle={Strain distribution in zincblende and wurtzite GaAs nanowires bent by a one-sided (In, Al)As shell: consequences for torsion, chirality, and piezoelectricity}, colorlinks,citecolor=blue,linkcolor=blue,urlcolor=blue}
    
\else\ifwias

	\documentclass[a4paper,12pt,twoside,english]{article}
	\usepackage{wiaspreprint}

	\pagefootleft{DOI 10.20347/WIAS.PREPRINT.3141}

	\usepackage{amsthm,amsmath,thmtools}
	\usepackage[utf8]{inputenc} 
	\usepackage[T1]{fontenc}
	\usepackage{lmodern}
	\usepackage{cleveref} 
	\usepackage[square,sort&compress,comma,numbers]{natbib}

	\definecolor{Brown}{rgb}{0.59, 0.29, 0.0}
	\author[Y. Hadjimichael, O. Brandt, C. Merdon, C. Manganelli, P. Farrell]{%
	Yiannis Hadjimichael\footnote{Weierstrass Institute \\
		Mohrenstra{\ss}e 39 \\ 10117 Berlin \\ Germany \\
		E-Mail: yiannis.hadjimichael@wias-berlin.de \\ 
		\hphantom{E-Mail:} christian.merdon@wias-berlin.de \\
		\hphantom{E-Mail:} patricio.farrell@wias-berlin.de
		},
	Oliver Brandt\footnote{Paul Drude Institute for  \\ Solid State Electronics (PDI) \\
		Hausvogteiplatz 5--7 \\ 10117 Berlin \\ Germany \\
		E-Mail: oliver.brandt@pdi-berlin.de
		},
	Christian Merdon\addmark{\hspace{5pt}}{1}, \\
	Costanza Manganelli\footnote{Leibniz Institute for \\ High Performance Microelectronics (IHP) \\
		Technologiepark 25 \\ 15236 Frankfurt (Oder) \\ Germany \\
		E-Mail: manganelli@ihp-microelectronics.com
		},
	Patricio Farrell\addmark{\hspace{5pt}}{1}\nofnmark{}
	}
	\title[Strain distribution in zincblende and wurtzite nanowires]
		{Strain distribution in zincblende and wurtzite GaAs nanowires bent by a one-sided (In, Al)As
		shell: Consequences for torsion, chirality, and piezoelectricity}
	\nopreprint{3141}	
	\nopreyear{2024}	
	\thanks{This work is supported by the Leibniz competition 2020 (NUMSEMIC, J89/2019) and the
		Deutsche Forschungsgemeinschaft (DFG) (project number: 434114264).
		We thank Oliver Marquardt and Taseer Anjum (Universität Siegen) for valuable discussions, and Vladimir
		Kaganer for a critical reading of the manuscript}

\else\ifarxiv

	\documentclass[12pt,a4paper,oneside]{article}

	\usepackage{amsmath,amsthm}	
	\usepackage{thmtools}
	\usepackage[top=3cm,bottom=3cm ,left=2cm,right=2cm,
		marginparsep=0.1cm,marginparwidth=2cm,footskip=2.0cm]{geometry}
	\usepackage{microtype} 
	\usepackage[pdftex]{graphicx} 
	\usepackage[dvipsnames]{xcolor}
	\usepackage[T1]{fontenc}	
	\usepackage[utf8]{inputenc} 
	\usepackage{newpxtext,newpxmath} 
	\usepackage[english]{isodate} 
	\usepackage[
		style=numeric-comp,
		hyperref=true,
		backend=biber,
		doi=false,
		url=false,
		isbn=false,
		date=year,
		backref=false,
		giveninits=true,
		sorting=nyt,
		maxcitenames=3,
		maxbibnames=100,
		block=none]{biblatex}
	\usepackage[
		linktocpage=true,
		colorlinks=true,
		linkcolor=blue,
		citecolor=red,
		urlcolor=blue,
		bookmarks=false]{hyperref}
	\RequirePackage[capitalize,nameinlink]{cleveref}[0.19]	

	\theoremstyle{plain}								
	
	\makeatletter
	\newcommand{\refcheckize}[1]{%
	  \expandafter\let\csname @@\string#1\endcsname#1%
	  \expandafter\DeclareRobustCommand\csname relax\string#1\endcsname[1]{%
	    \csname @@\string#1\endcsname{##1}\@for\@temp:=##1\do{\wrtusdrf{\@temp}\wrtusdrf{{\@temp}}}}%
	  \expandafter\let\expandafter#1\csname relax\string#1\endcsname
	}
	\newcommand{\refcheckizetwo}[1]{%
	  \expandafter\let\csname @@\string#1\endcsname#1%
	  \expandafter\DeclareRobustCommand\csname relax\string#1\endcsname[2]{%
	    \csname @@\string#1\endcsname{##1}{##2}\wrtusdrf{##1}\wrtusdrf{{##1}}\wrtusdrf{##2}\wrtusdrf{{##2}}}
	  \expandafter\let\expandafter#1\csname relax\string#1\endcsname
	}
	\makeatother

	\refcheckize{\Cref}
	\refcheckizetwo{\crefrange}
	\refcheckizetwo{\Crefrange}

	\crefname{section}{section}{sections}
	\crefname{subsection}{section}{subsections}
	\Crefname{subsection}{Section}{Subsections}
	
	\DeclareNameAlias{default}{family-given}
	\newbibmacro*{string+doiurlisbn}[1]{
		\iffieldundef{doi}{
			\iffieldundef{url}{
				\iffieldundef{isbn}{
					\iffieldundef{issn}{#1}
						{\href{http://books.google.com/books?vid=ISSN\thefield{issn}}{#1}}
				}
						{\href{http://books.google.com/books?vid=ISBN\thefield{isbn}}{#1}}
			}
			{\href{\thefield{url}}{#1}}
		}
		{\href{http://dx.doi.org/\thefield{doi}}{#1}}
	}
	\DeclareFieldFormat*{title}{\usebibmacro{string+doiurlisbn}{#1\isdot}}
	\renewcommand*{\newunitpunct}{\addcomma\space}
	\DeclareFieldFormat{journaltitle}{\mkbibitalic{#1}\newunitpunct}
	\DeclareFieldFormat{booktitle}{\mkbibitalic{#1}}
	\renewbibmacro{in:}{}
	\AtEveryBibitem{\clearfield{number}}
	\DeclareSourcemap{
		\maps[datatype=bibtex]{
			\map[overwrite=true]{
				\step[fieldsource=fjournal]
				\step[fieldset=journal, origfieldval]
			}
		}
	}
	\title{Strain distribution in zincblende and wurtzite GaAs nanowires bent by a one-sided (In, Al)As shell:
		consequences for torsion, chirality, and piezoelectricity}
	\author{Yiannis Hadjimichael\thanks{WIAS---Weierstraß-Institut für Angewandte Analysis und Stochastik
        (Weierstrass Institute for Applied Analysis and Stochastics), Mohrenstra{\ss}e 39, 10117 Berlin,
        Germany,
	\newline
	(emails: \texttt{hadjimichael@wias-berlin.de}, \texttt{merdon@wias-berlin.de},
		\texttt{farrell@wias-berlin.de}).}
	\and
	Oliver Brandt\thanks{PDI---Paul-Drude-Institut für Festkörperelektronik
		(Paul Drude Institute for Solid State Electronics), 
		Hausvogteiplatz 5--7, 10117 Berlin, Germany, (email: \texttt{brandt@pdi-berlin.de}).}
	\and
	Christian Merdon\samethanks[1]
	\and
	Costanza Manganelli\thanks{IHP---Leibniz-Institut für innovative Mikroelektronik (Leibniz Institute for
		High Performance Microelectronics), Technologiepark 25, 15236 Frankfurt (Oder), Germany,
        (email: \texttt{manganelli@ihp-microelectronics.com}).}
	\and
	Patricio Farrell\samethanks[1]
	}

\fi\fi\fi

\usepackage{mathtools}
\usepackage{booktabs} 
\usepackage{bm}
\usepackage{siunitx}
\usepackage{array}
\usepackage{multirow}
\usepackage{marginnote}
\usepackage{ulem}
\usepackage{threeparttable}
\usepackage{soul}
\usepackage[mathlines]{lineno} 
\usepackage[colorinlistoftodos,prependcaption,textsize=scriptsize,color=lightgray]{todonotes}

\newcommand*\samethanks[1][\value{footnote}]{\footnotemark[#1]} 	
\newcommand{\IR}{\mathbb{R}}  									
\newcommand{\transpose}[1]{#1^{\raisemath{0.5pt}{\mathsf{T}}}}	
\newcommand{\invtranspose}[1]{#1^{-\mathsf{T}}}					
\newcommand{\elstrain}{\mathsfit{\varepsilon}}
\newcommand{\totstrain}{\mathsfit{\epsilon}}

\newcommand{\stiff}{\bm{\mathsfit{C}}}
\newcommand{\F}{\bm{\mathsfit{F}}}
\newcommand{\I}{\bm{\mathsfit{I}}}
\newcommand{\M}{\bm{\mathsfit{M}}}
\DeclareSIUnit\angstrom{\text{Å}}											

\makeatletter
	\ifjournal
		\newcommand{\refcite}[1]{Ref.~[\onlinecite{#1}]}
		\newcommand{\mycite}[2]{#1\cite{#2}}
	\else
		\newcommand{\refcite}[1]{\cite{#1}}
		\newcommand{\mycite}[2]{ \cite{#2}#1}
	\fi
\makeatother

\makeatletter
	\newcommand{\raisemath}[1]{\mathpalette{\raisem@th{#1}}}
	\newcommand{\raisem@th}[3]{\raisebox{#1}{$#2#3$}}
\makeatother

\makeatletter
\renewcommand*\env@matrix[1][\arraystretch]{%
	\edef\arraystretch{#1}%
	\hskip -\arraycolsep
	\let\@ifnextchar\new@ifnextchar
	\array{*\c@MaxMatrixCols c}}
\makeatother

\newcount\todocount

\newcommand{\checkxpos}[3][]{%
  \ifdim \zposx{#2}sp < 20000000sp%
    \mynote[#1]{#3}%
  \else%
    \note[#1]{#3}%
  \fi%
}

\newcommand{\mynote}[2][]{{%
  \let\marginpar\marginnote
  \reversemarginpar
  \renewcommand{\baselinestretch}{0.8}%
  \todo[#1]{#2}}}

\newcommand{\note}[2][]{\renewcommand{\baselinestretch}{0.8}\todo[#1]{#2}}

\DeclareFontFamily{OMX}{lmex}{}
\DeclareFontShape{OMX}{lmex}{m}{n}{<-> lmex10}{}

\ifarxiv
	\DeclareMathAlphabet{\mathsfit}{T1}{\sfdefault}{\mddefault}{\itdefault}
	\SetMathAlphabet{\mathsfit}{bold}{T1}{\sfdefault}{\bfdefault}{\itdefault}
\else
	\DeclareMathAlphabet{\mathsfit}{T1}{\sfdefault}{\mddefault}{\sldefault}
	\SetMathAlphabet{\mathsfit}{bold}{T1}{\sfdefault}{\bfdefault}{\sldefault}
\fi

\newcolumntype{P}[1]{>{\centering\arraybackslash}p{#1}}
\newcolumntype{M}[1]{>{\centering\arraybackslash}m{#1}}

\graphicspath{{figures/}, {figures_supplementary_material/},{cover_letter/wiascd/}}

\begin{document}

\ifjournal
	\title{Strain distribution in zincblende and wurtzite GaAs nanowires bent by a one-sided (In, Al)As stressor
		shell: consequences for torsion, chirality, and piezoelectricity}
	\author{Yiannis Hadjimichael}
	\email[Electronic email: ]{yiannis.hadjimichael@wias-berlin.de}
	\affiliation{WIAS---Weierstraß-Institut für Angewandte Analysis und Stochastik
		(Weierstrass Institute for Applied Analysis \\ and Stochastics),
		Mohrenstra{\ss}e 39, 10117 Berlin, Germany}
	\author{Oliver Brandt}
	\affiliation{PDI---Paul-Drude-Institut für Festkörperelektronik
		(Paul Drude Institute for Solid State Electronics), 
		Hausvogteiplatz 5--7, 10117 Berlin, Germany}
	\author{Christian Merdon}
	\affiliation{WIAS---Weierstraß-Institut für Angewandte Analysis und Stochastik
		(Weierstrass Institute for Applied Analysis \\ and Stochastics),
		Mohrenstra{\ss}e 39, 10117 Berlin, Germany}
	\author{Costanza Manganelli}
	\affiliation{IHP---Leibniz-Institut für innovative Mikroelektronik
		(Leibniz Institute for High Performance Microelectronics),
		Technologiepark 25, 15236 Frankfurt (Oder), Germany}
	\author{Patricio Farrell}
	\affiliation{WIAS---Weierstraß-Institut für Angewandte Analysis und Stochastik
		(Weierstrass Institute for Applied Analysis \\ and Stochastics),
		Mohrenstra{\ss}e 39, 10117 Berlin, Germany}
	\linenumbers\relax
\else
	\maketitle
\fi

\begin{abstract}
We present a finite-strain model that is capable of describing the large deformations in bent nanowire heterostructures. The model incorporates a nonlinear strain formulation derived from the first Piola-Kirchhoff stress tensor, coupled with an energy functional that effectively captures the lattice-mismatch-induced strain field. We use the finite element method to solve the resulting partial differential equations and extract cross-sectional maps of the full strain tensor for both zincblende and wurtzite nanowires with lattice-mismatched core and one-sided stressor shell. In either case, we show that the bending is essentially exclusively determined by $\elstrain_{zz}$. However, the distinct difference in shear strain has important consequences with regard to both the mechanical deformation and the existence of transverse piezoelectric fields in the nanowires.
\end{abstract}
\ifjournal
    \maketitle
\else
	\section{Introduction}\label{sec:introduction}
\fi
Strain engineering is a powerful means for manipulating and tuning of the electronic band structure of semiconductors\mycite{.}{oreilly_Semicond.Sci.Technol._1989} In planar heterostructures, combining layers with different lattice constants generates a spatially uniform biaxial strain that strongly modifies the energy and dispersion of the electronic bands in the strained layer, and which is routinely used for various electronic and optoelectronic applications of semiconductors\mycite{.}{oreilly_IEEEJ.QuantumElectron._1994,maiti__2001,
adams_IEEEJ.Sel.Top.QuantumElectron._2011,schriever_Materials_2012,dai_Adv.Mater._2019}

However, strain \textit{gradient} engineering, i.\,e., the use of the effects of spatially nonuniform strain, has thus far remained largely unexplored, and has only recently emerged as one of the key topics for future solid-state research. In fact, sufficiently large strain gradients in inorganic materials unlock new physical properties such as, most prominently, flexoelectricity\mycite{,}{zubko_Annu.Rev.Mater.Res._2013,yudin_Nanotechnology_2013,wang_Prog.Mater.Sci._2019,xia_ACSAppl.Mater.Interfaces_2024} but may also induce ferroelectricity, superconductivity, and flexomagnetism\mycite{.}{du_Appl.Phys.Lett._2023} Continuously tunable strain gradients can furthermore be utilized to realize spectrally indistinguishable quantum emitters\mycite{.}{maity_Phys.Rev.Appl._2018} While the magnitude of such gradients in inorganic bulk crystals is limited by their low fracture strength, free-standing nanostructures such as nanomembranes or nanowires are tolerant to loads close to the elastic limit of the respective material\mycite{,}{stan_NanoLett._2012,zhang_Sci.Adv._2016} and can thus sustain strain gradients in excess of $2\times 10^{-3}$/nm.

The flexibility and quasi-one-dimensional nature of nanowires offer new degrees of freedom for novel heterostructure design and strain engineering\mycite{.}{boxberg_Adv.Mater._2012} Bent nanowires have therefore attracted attention in both the semiconductor and the piezotronics fields. For example, dispersed ZnO microwires have been bent mechanically by micromanipulation to induce charge carrier drift in the strain (and thus bandgap) gradient across the wire's cross section\mycite{.}{dietrich_Appl.Phys.Lett._2011,xu_NanoLett._2012,fu_ACSNano_2014,fu_Adv.Mater._2014} The bending of free-standing ZnO nanowires by an external load such as the tip of an atomic force microscope was theoretically investigated by several groups to compute the resulting piezo- and flexoelectric fields\mycite{.}{gao_NanoLett._2007,wang_Adv.Mater._2007,gao_NanoLett._2009,liu_SmartMater.Struct._2012,zhang_J.Appl.Phys._2016,wang_Nanotechnology_2018}

A novel method to uniformly bend all nanowires in an ordered array has been developed by researchers from the Paul-Drude-Institute\mycite{}{lewis_NanoLett._2018,davtyan_J.Appl.Crystallogr._2020} and McMaster University\mycite{.}{mcdermott_ACSAppl.NanoMater._2021,mcdermott_ACSAppl.NanoMater._2024} By exploiting the directional nature of the impinging atom fluxes in molecular beam epitaxy, they fabricated nanowires that were bent in a controlled fashion by highly lattice-mismatched stressor shells deposited on a single side of the nanowires. This approach allows precise control over the degree of bending and the resulting strain gradients, eliminates the need for external mechanical micromanipulation, and produces arrays of free-standing nanowires with deterministic positioning and uniform bending. However, the current understanding of the strain distribution in these nanowires is rudimentary in the sense that it neglects all strain components except $\elstrain_{zz}$\mycite{.}{dietrich_Appl.Phys.Lett._2011,xu_NanoLett._2012,fu_ACSNano_2014,fu_Adv.Mater._2014,lewis_NanoLett._2018}

In this paper, we use a three-dimensional finite-strain model based on the Green-Lagrange total strain tensor, including geometric nonlinearity, to describe the large deformations induced by the lattice mismatch between the nanowire core and a one-sided lattice-mismatched stressor shell\mycite{.}{ciarlet__2021,hadjimichael_Int.J.Numer.MethodsEng._2024} We explicitly consider GaAs/(In,Al)As core/shell nanowire heterostructures as realized in \refcite{lewis_NanoLett._2018}. These nanowires consist of a GaAs core with a regular hexagonal cross-section and a one-sided (In,Al)As stressor shell that bends the core. We derive the full strain tensor for both zincblende (ZB) and wurtzite (WZ) nanowires, each for shells deposited in the two perpendicular crystallographic directions, and compare the numerical results to approximate analytical results. Furthermore, we point out important differences between the two polytypes and their implications for the existence of transverse piezoelectric fields in the nanowires.

\ifjournal
	\begin{figure*}
\else
	\section{Finite-strain model}\label{sec:model}
	\begin{figure}[t!]
\fi	
	\centering
	\includegraphics[width=0.49\textwidth]{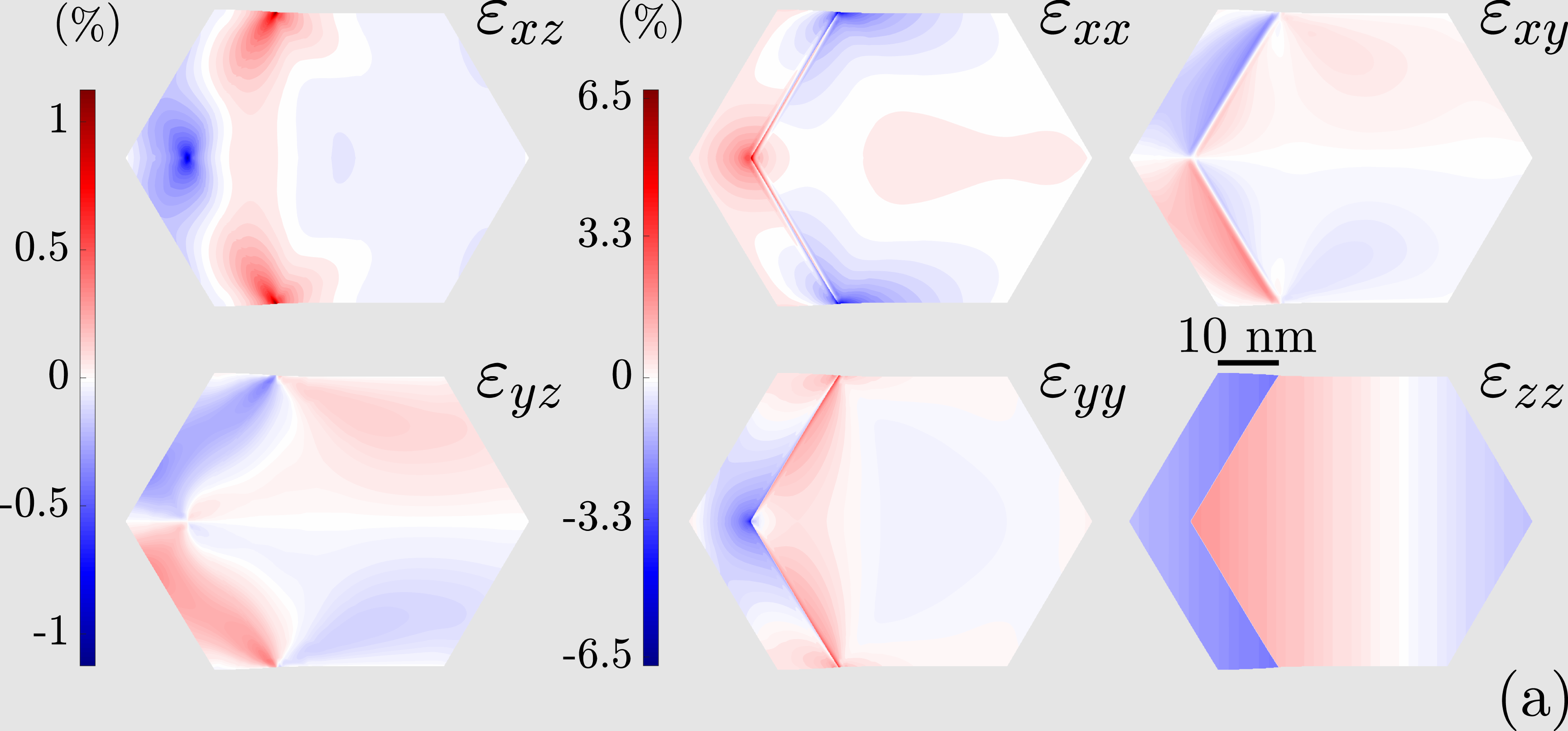}\hfill
	\includegraphics[width=0.49\textwidth]{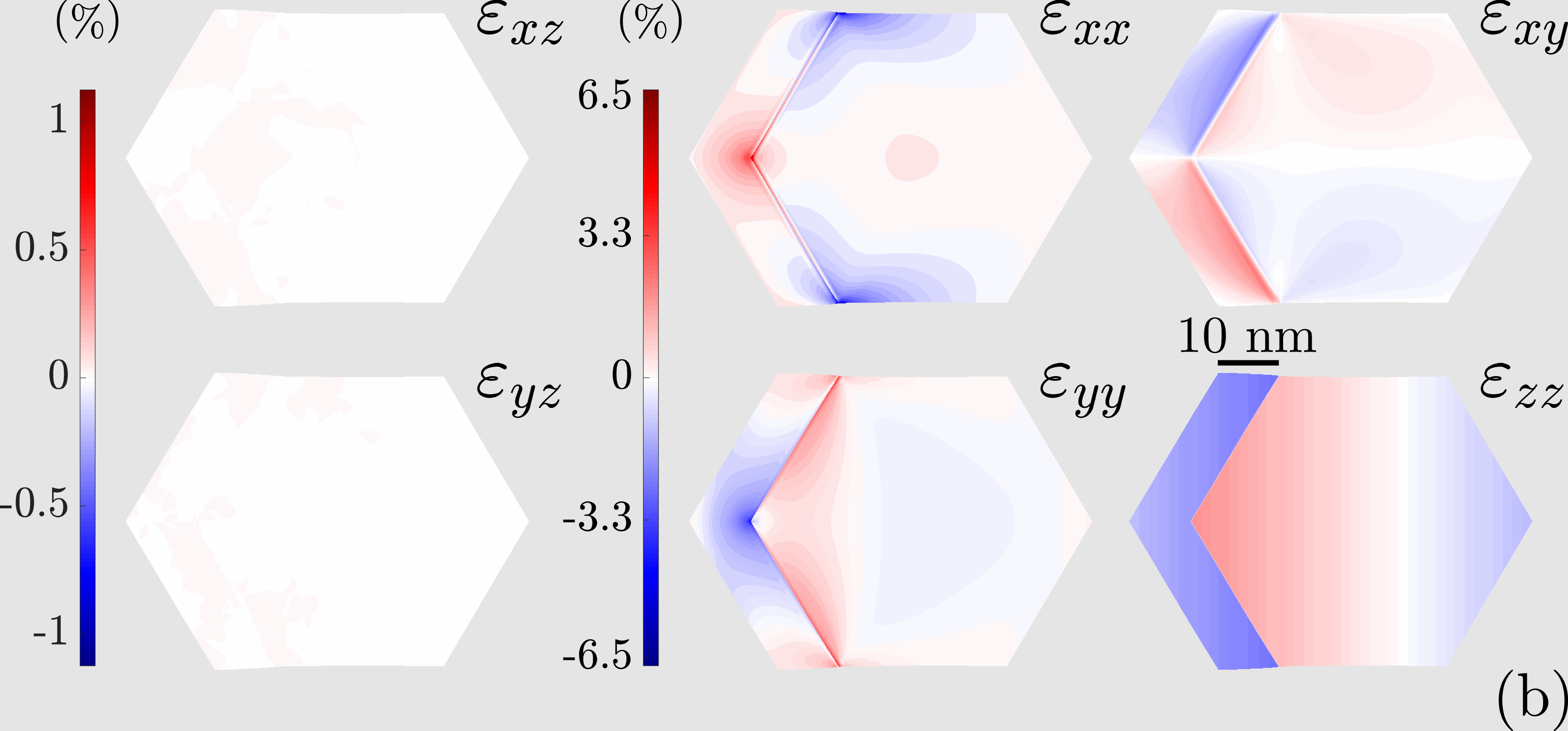}\\[10pt]
	\includegraphics[width=0.49\textwidth]{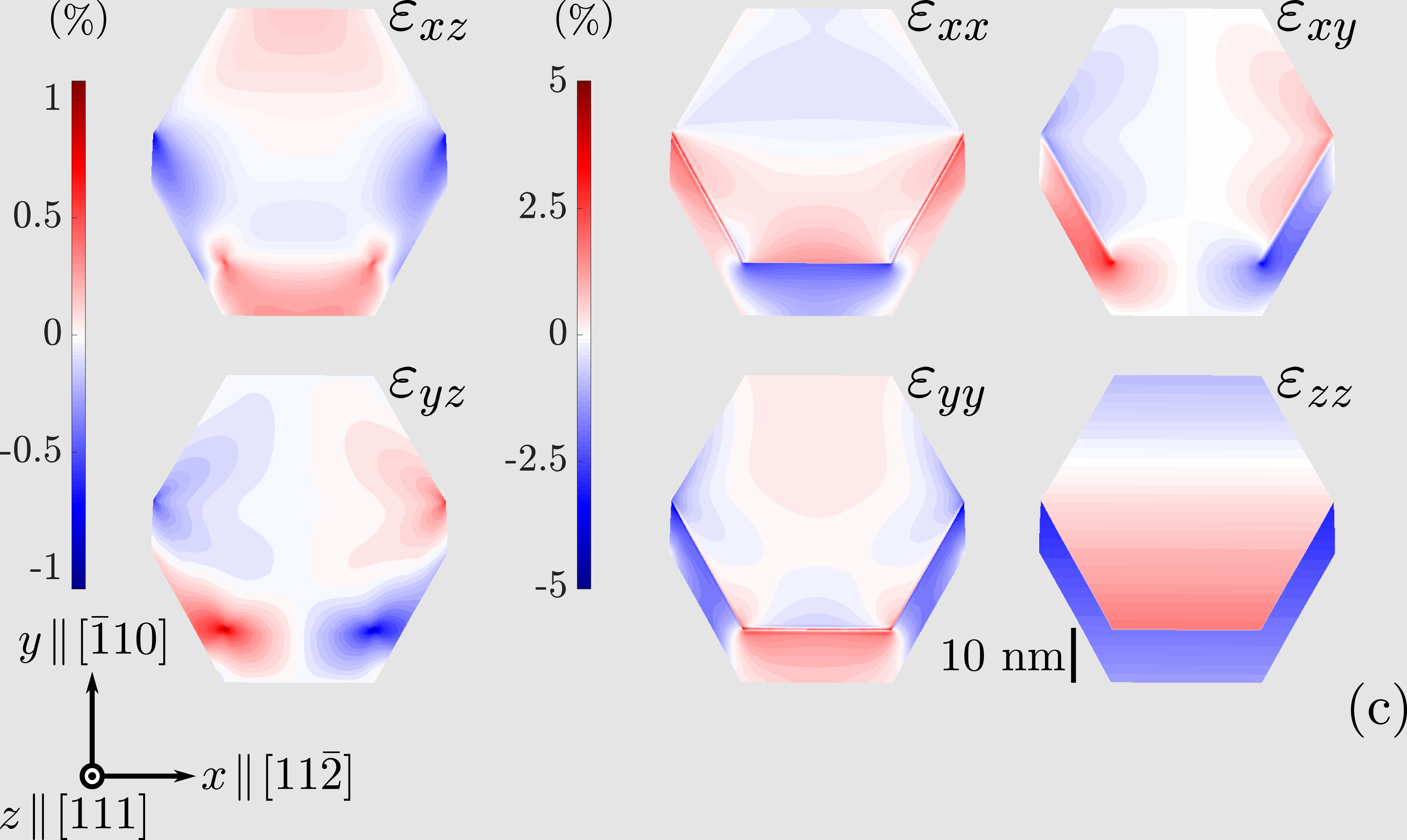}\hfill
	\includegraphics[width=0.49\textwidth]{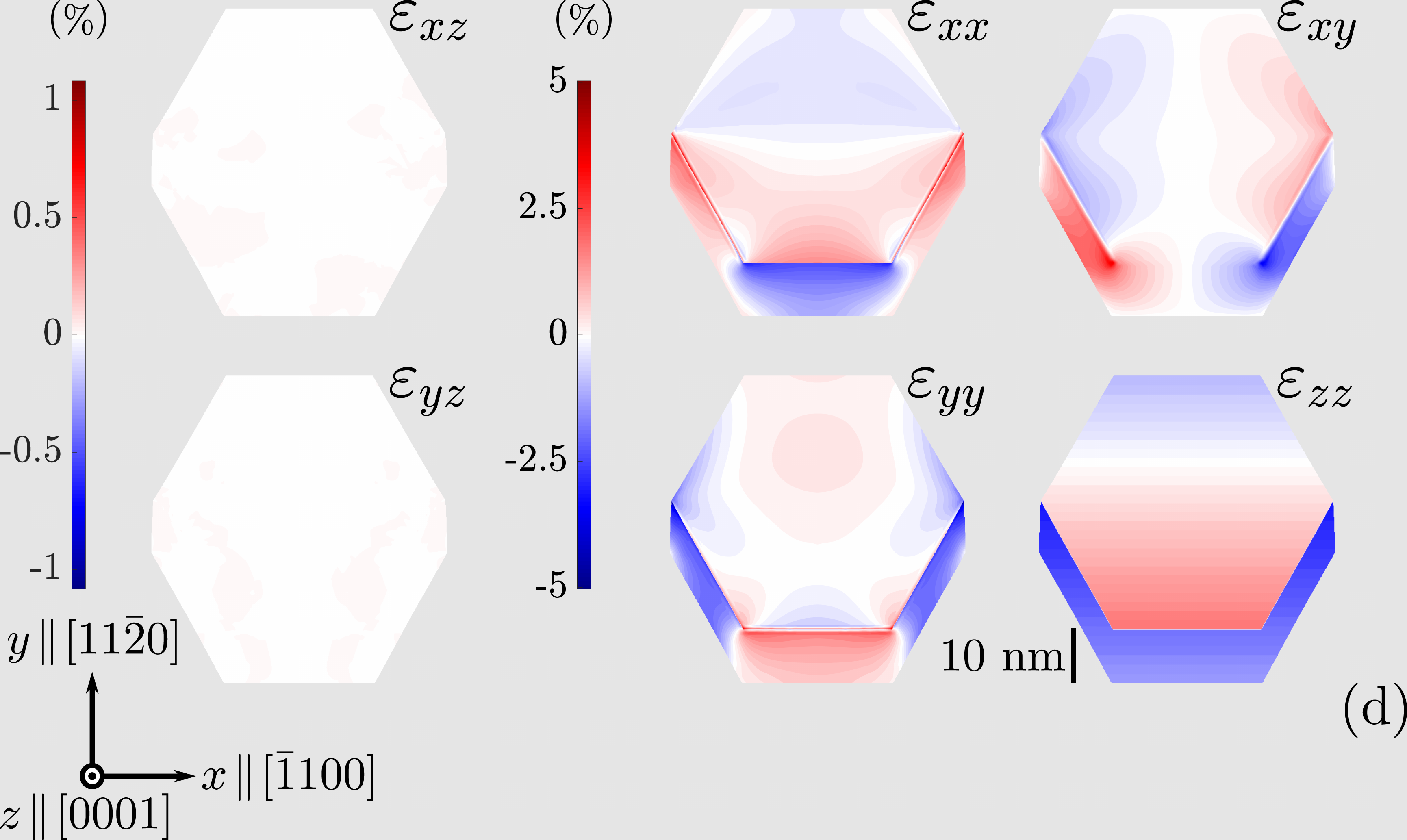}
	\caption{Elastic strain components computed on the central cross section of a 2~\textmu m long GaAs/$\text{In}_{0.5}\text{Al}_{0.5}\text{As}$ core/shell nanowire, using two different cross-sectional geometries. In the top panel, the shell is applied on a vertex of the hexagonal core: (a) along the $[11\bar{2}]$ direction for ZB-[111] nanowires, and (b) along the $[\bar{1}100]$ direction for WZ-[0001] nanowires. In the bottom panel, the shell is applied on the edge of the hexagonal core, i.\,e., in (c), along the $[\bar{1}10]$ direction for ZB-[111] nanowires, and in (d) along the $[11\bar{2}0]$ direction for WZ-[0001] nanowires. In each case, we separate the $\elstrain_{xz}$ and $\elstrain_{yz}$ shear strain components since their values are smaller than the normal components. In the ZB-111 case [see (a) and (c)], the $\elstrain_{xz}$ and $\elstrain_{yz}$ components are approximately five to six times smaller than those of the normal strain, but in the WZ-[0001] case [see (b) and (d)], they are zero\ifjournal\mycite{.}{Note1}\else .\fi}
	\label{fig:ZB_WZstrain}
\ifjournal
	\end{figure*}
\else
	\end{figure}
\fi	

Our model is numerically solved in the reference (Lagrangian) configuration and is based on a predeformation tensor
$\M$ which takes into account the influence of the lattice mismatch between the two materials (see the Julia package \texttt{StrainedBandstructures.jl}\mycite{.}{hadjimichael_Github_SBSP} Hence, $\M$ differs in each region of the composite structure. Let $\bm{u}\in \mathbb{R}^3$ denote the displacement, $\F \coloneqq \nabla\bm{\phi} = \I +\bm{\nabla u}$ the deformation gradient tensor, and $\stiff \in \IR^{3\times 3 \times 3 \times 3}$ the fourth-rank stiffness tensor. In our model, we use a linear stress-strain relation (Hooke's law) and consider the Green-Lagrange total strain tensor with a geometric nonlinearity, i.e.\, $\bm{\totstrain}(\F) \coloneqq \frac{1}{2}(\transpose{\F}\F - \I) = \frac{1}{2}\left(\nabla \bm{u} + \transpose{\nabla \bm{u} } + \transpose{\nabla \bm{u} }\nabla \bm{u} \right)$. Then, our model can be expressed by the partial differential equation\mycite{}{hadjimichael_Int.J.Numer.MethodsEng._2024}
\begin{align}\label{eq:divergence_energy_equilibrium}
	-\operatorname{div}\left(\operatorname{det}\bigl(\M\bigr)\F(\bm{u})\M^{-1}
		\bigl(\stiff : \bm{\elstrain}(\bm{u})\bigr)\invtranspose{\M}\right) = \bm{0},
\end{align}
which is solved for the displacement $\bm{u}$. In this equation, $\bm{\elstrain} = \invtranspose{\M}\bm{\totstrain}(\F)\M^{-1} - \frac{1}{2}(\I-\invtranspose{\M}\M^{-1})$ is the elastic strain. The model is supplied with homogeneous Neumann boundary conditions at all facets except at the interface between the nanowire and the substrate, where we apply Dirichlet boundary conditions.

\ifjournal\else
	\section{Strain profiles in bent ZB/WZ nanowires}\label{sec:strain_ZBWB}
\fi
Since none of the strain components change along the length of the nanowire (except in close vicinity to the nanowire ends), the problem can be reduced to two dimensions spanned by the cross-sectional coordinates perpendicular to the bending direction. Figure~\ref{fig:ZB_WZstrain} shows the six elastic strain components in the center cross-section (taken at 1~\textmu m) of a 2~\textmu m long GaAs/$\text{In}_{0.5}\text{Al}_{0.5}\text{As}$ core/shell nanowire with a core diameter $d = 2 s = 100/\sqrt{3}$\,nm, where $s$ denotes both the circumradius and the side length of the regular hexagon forming the nanowire core, and a shell thickness $\delta =10$\,nm\ifjournal\mycite{.}{lewis_NanoLett._2018_SI}\else .\footnote{See the supporting information of \refcite{lewis_NanoLett._2018}.}\fi\
We consider two different polytypes [ZB in (a) and (c), WZ in (b) and (d)] and two different core/shell geometries (the two top and bottom rows). The curvature $\kappa$ of these four different bent nanowires is only slightly different and corresponds roughly to 0.52\,\textmu m$^{-1}$. As an illustration,
\ifjournal
	Fig.~S1 in the supplementary material
\else
	Fig.~\ref{fig:Appx_nanowire3D} in the appendix
\fi
provides a three-dimensional view of the ZB-[111] nanowire from Fig.~\ref{fig:ZB_WZstrain}(c).

In all four cases, the normal strain components $\elstrain_{xx}$, $\elstrain_{yy}$, and $\elstrain_{zz}$ are large and very similar in magnitude for both polytypes and crystallographic bending directions. The same applies to the in-plane shear strain component $\elstrain_{xy}$.
However, the values of the two out-of-plane shear strain components $\elstrain_{xz}$ and $\elstrain_{yz}$ differ drastically between the two crystal structures. These components have the same order of magnitude as the normal components for ZB-[111] nanowires [see Figs.~\ref{fig:ZB_WZstrain}(a) and \ref{fig:ZB_WZstrain}(c)], but are actually zero for WZ-[0001] nanowires [see Figs.~\ref{fig:ZB_WZstrain}(b) and \ref{fig:ZB_WZstrain}(d)].\footnote{The finite values on the order of $10^{-5}$ are a numerical artifact resulting from the comparatively coarse grid used for these computations. With increasing spatial refinement, these components asymptotically approach zero.}
The origin of this difference in magnitude of $\elstrain_{xz}$ and $\elstrain_{yz}$ lies in the structure of the stiffness tensors for the two different polytypes. For both direct comparison and numerical efficiency\mycite{,}{schulz_Phys.Rev.B_2011} we transform the stiffness matrix for the ZB structure (defined with the unit vectors along the $\langle 001 \rangle$ axes) into the nanowire coordinate system spanned by the $[111]$, $[\bar{1}10]$, and $[11\bar{2}]$ directions. No such transformation is necessary for the WZ structure, whose unit vectors naturally align with the nanowire coordinates.

In Voigt notation, the stiffness matrices $\bm{C}$ for WZ-[0001]- and ZB-[111]-oriented nanowires are given by\mycite{:}{schulz_Phys.Rev.B_2011}
\begin{align*}
\bm{C}^{0001} &\coloneqq \begin{pmatrix}[1.2]
						C_{11}^\text{\tiny WZ} & C_{12}^\text{\tiny WZ} & C_{13}^\text{\tiny WZ} & 0 & 0 & 0 \\
						C_{12}^\text{\tiny WZ} & C_{11}^\text{\tiny WZ} & C_{13}^\text{\tiny WZ} & 0 & 0 & 0 \\
						C_{13}^\text{\tiny WZ} & C_{13}^\text{\tiny WZ} & C_{33}^\text{\tiny WZ} & 0 & 0 & 0 \\
						0 & 0 & 0 & C_{44}^\text{\tiny WZ}	& 0 &0 \\
						0 & 0 & 0 & 0 & C_{44}^\text{\tiny WZ} & 0 \\
						0 & 0 & 0 & 0 & 0 & C_{66}^\text{\tiny WZ}
					\end{pmatrix}, &
\end{align*}
and
\begin{align*}
\bm{C}^{111} &\coloneqq \begin{pmatrix}[1.2]
						C_{11}^\text{\tiny ZB} & C_{12}^\text{\tiny ZB} & C_{13}^\text{\tiny ZB} & 0 & C_{15}^\text{\tiny ZB} & 0 \\
						C_{12}^\text{\tiny ZB} & C_{11}^\text{\tiny ZB} & C_{13}^\text{\tiny ZB} & 0 & -C_{15}^\text{\tiny ZB} & 0 \\
						C_{13}^\text{\tiny ZB} & C_{13}^\text{\tiny ZB} & C_{33}^\text{\tiny ZB} & 0 & 0 & 0 \\
						0 & 0 & 0 & C_{44}^\text{\tiny ZB}	& 0 &-C_{15}^\text{\tiny ZB} \\
						C_{15}^\text{\tiny ZB} & -C_{15}^\text{\tiny ZB} & 0 & 0 & C_{44}^\text{\tiny ZB} & 0 \\
						0 & 0 & 0 & -C_{15}^\text{\tiny ZB} & 0 & C_{66}^\text{\tiny ZB}
					\end{pmatrix}. &
\end{align*}
Since there is no complete set of parameters available for the WZ materials, particularly for InAs and AlAs, we
use a linear transformation (quasi-cubic approximation) to obtain $C_{ij}^\text{\tiny WZ}$ from the stiffness
constants of the corresponding ZB materials in their natural ($\langle 001 \rangle$) basis
\mycite{.}{martin_Phys.Rev.B_1972}
Likewise, the relations between the $C_{ij}^\text{\tiny ZB}$ in the rotated basis ($\langle 111 \rangle$) and
those in the natural coordinate system ($\langle 001 \rangle$) of the ZB crystal are given in 
\refcite{schulz_Phys.Rev.B_2011}.
For this particular orientation the components of the stiffness matrices are identical
($C_{ij}^\text{\tiny WZ} =  C_{ij}^\text{\tiny ZB}$), except for the additional component $C_{15}$ in
$\bm{C}^{111}$.
The values for the lattice, stiffness, and piezoelectric constants used in the present work for both ZB and WZ GaAs and (In,Al)As are given in
\ifjournal
	Tab.~S1 in the supplementary material.
\else
	Tab.~\ref{tab:Appx_parameters} in the appendix.
\fi

The key difference between $\bm{C}^{111}$ and $\bm{C}^{0001}$ is the element $C_{15}$, which is zero for the WZ but non-zero for the ZB case. For ZB-GaAs, for example, we obtain
a value of 12.85\,GPa. The structure of these matrices properly reflects the trigonal ($C_{3v}$) and hexagonal ($C_{6v}$) symmetry of ZB-[111]- and WZ-[0001] nanowires. Note that different but equivalent representations of $\bm{C}^{111}$ can be obtained depending on the exact choice of the coordinate transformation\mycite{.}{cowin_RationalContinuaClassicalandNew_2003,boxberg_Adv.Mater._2012}

\ifjournal
	\begin{figure}[t!]
\else
	\section{Torsion}\label{sec:torsion}
	\begin{figure}[b!]
\fi
	\centering
	\ifjournal
		\includegraphics[width=0.494\columnwidth]{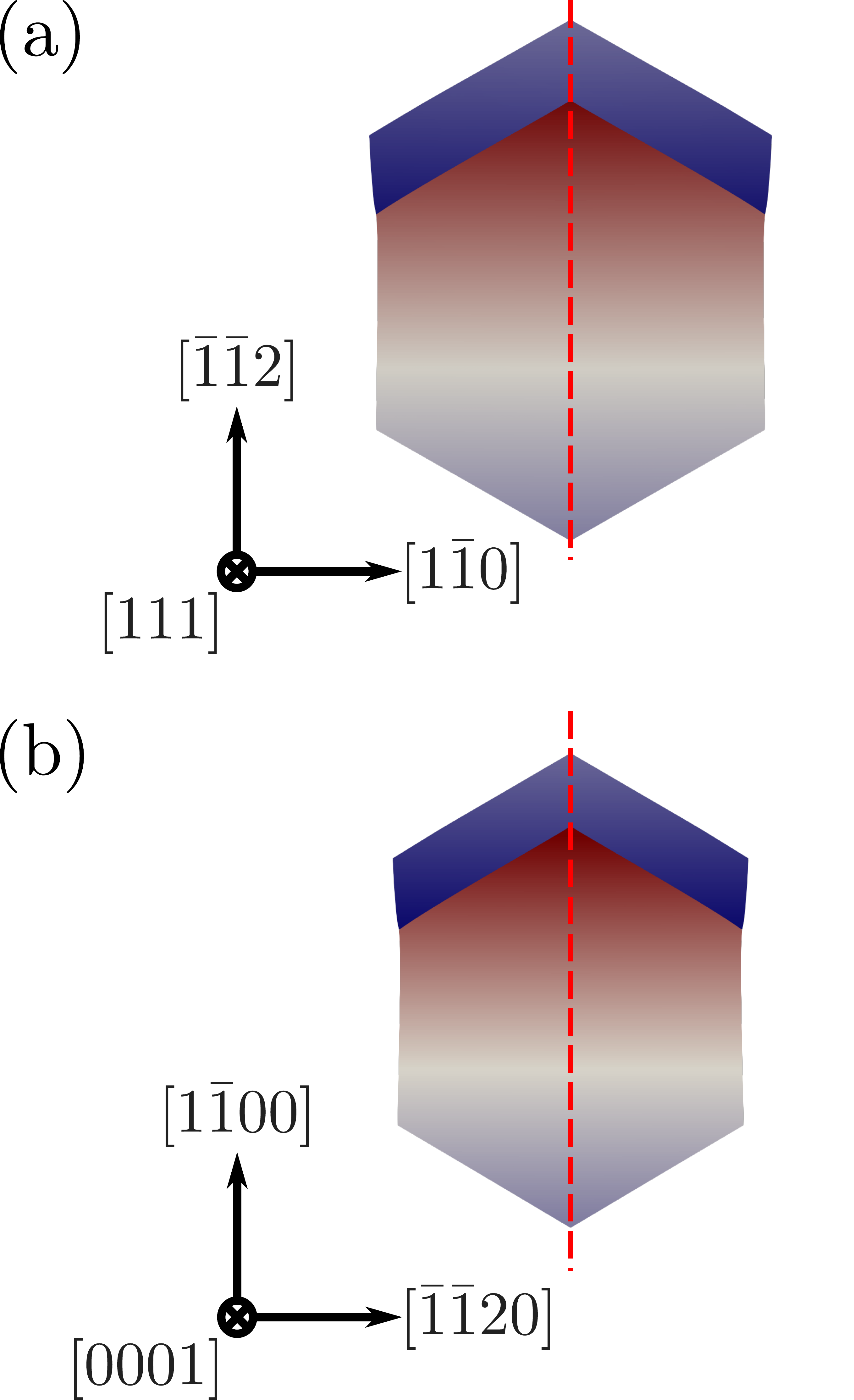}
		\includegraphics[width=0.494\columnwidth]{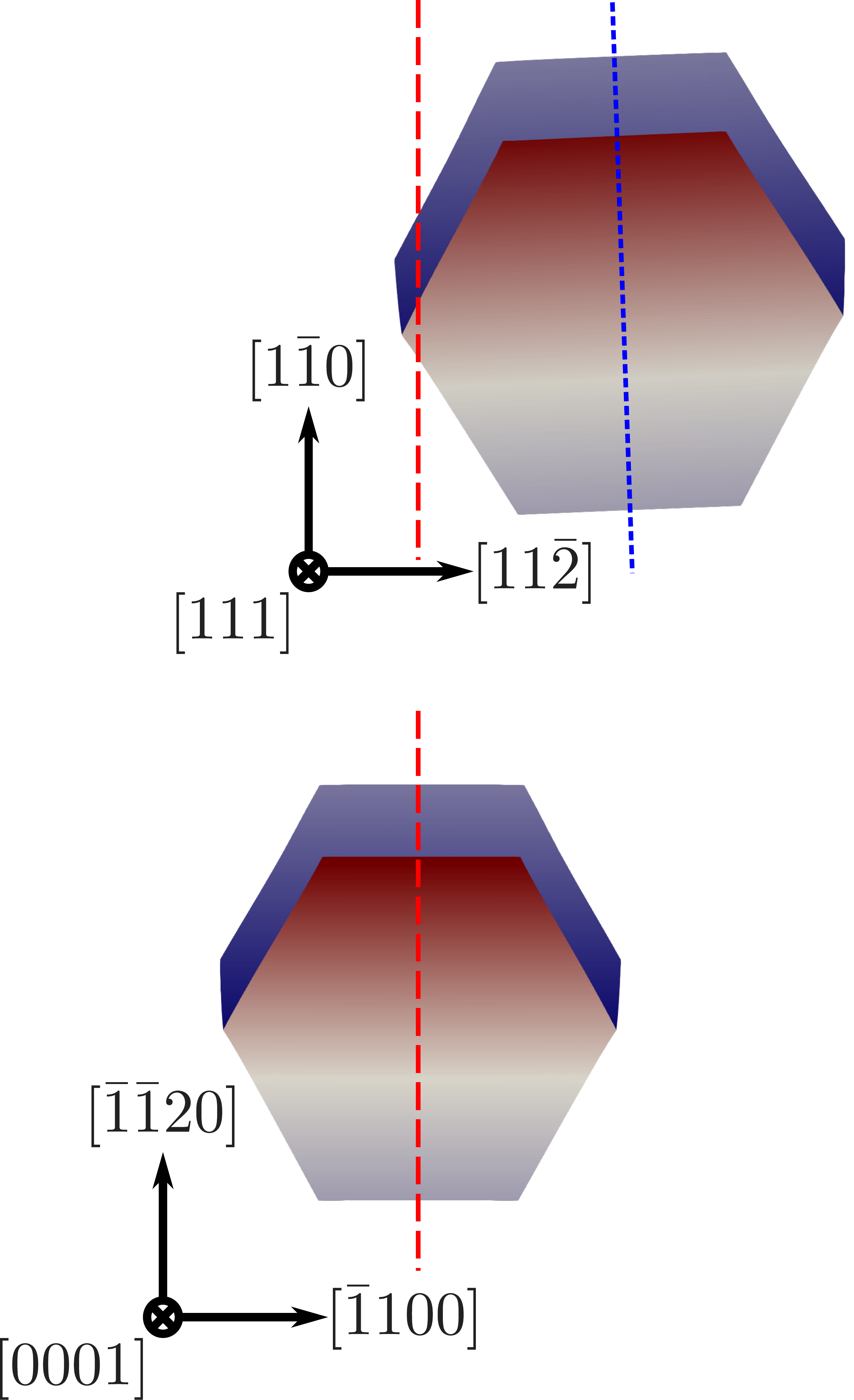}
	\else
		\includegraphics[width=0.35\columnwidth]{torsion1}\qquad\qquad
		\includegraphics[width=0.35\columnwidth]{torsion2}
	\fi
	\caption{Bent GaAs/InAs nanowires viewed along the normal of their top facet from the same viewpoint (inside the nanowire). Due to the different curvature and thus distance from the viewpoint, the nanowires seem to be of different size, but they all have an identical core diameter of $100/\sqrt{3}$\,nm. The color gradient depicts the variation of $\elstrain_{zz}$ analogously to Fig.~\ref{fig:ZB_WZstrain}. In (a), we show ZB-[111] nanowires with the stressor shell deposited along the (left) $[11\bar{2}]$ and (right) $[\bar{1}10]$ directions corresponding to Fig.~\ref{fig:ZB_WZstrain}(a) and \ref{fig:ZB_WZstrain}(c). In (b), the corresponding WZ-[0001] nanowires are displayed with the stressor shell deposited along the (left) $[\bar{1}100]$ and (right) $[11\bar{2}0]$ directions, corresponding to Fig.~\ref{fig:ZB_WZstrain}(b) and \ref{fig:ZB_WZstrain}(d). The red dashed line represents the bending plane for each of the four different cases. In three of these cases, the nanowire stays exactly in the bending plane. However, for ZB-[111] nanowires with the shell applied in the $[\bar{1}10]$ direction, the nanowires is laterally displaced and exhibits a torsion visualized by the blue dotted line.}
    \label{fig:torsion}
\end{figure}
An as unexpected as interesting consequence of this distinct difference in the strain distribution in ZB-[111]- and WZ-[0001]-oriented nanowires is displayed in Fig.~\ref{fig:torsion}, which views the nanowires along the normal of their top facet for our four different bent nanowire configurations. To maximize the displacements, we here consider GaAs nanowires with a pure InAs stressor shell, representing the maximum lattice mismatch available in this materials system. For the WZ-[0001] nanowires shown in Fig.~\ref{fig:torsion}(b), the nanowire bends strictly in the bending plane indicated by the red dashed lines. The same is true for the ZB-[111] nanowire shown in the left panel of Fig.~\ref{fig:torsion}(a) with the stressor shell applied along the $[11\bar{2}]$ direction. However, for the ZB-[111] nanowire shown in the right panel of Fig.~\ref{fig:torsion}(a) with the stressor shell applied along the $[\bar{1}10]$ direction, we observe a displacement of the nanowire out of the bending plane, and an anticlock-wise rotation about the nanowire's [111] axis. 

This torsion of the nanowire induces a left-handed helical chirality. At the first glance, this result may appear surprising, since the zincblende structure is cubic, and materials of cubic elastic symmetry are not chiral. However, as mentioned above, the rotated stiffness matrix of the ZB-[111]-oriented nanowires exhibits a trigonal elastic symmetry, which is in fact the highest elastic symmetry that permits chirality\mycite{.}{fraldi_J.Elast._2002} In contrast, the stiffness matrix of the WZ-[0001]-oriented nanowires belongs to the class of transverse isotropic materials, which are not chiral\mycite{.}{cowin__2013}

The torsion of the ZB-[111]-oriented nanowire is induced by the large out-of-plane shear strain components, which in turn originate from the finite value of the symmetry-breaking or chiral stiffness constant $C_{15}$\mycite{.}{fraldi_J.Elast._2002,cowin_RationalContinuaClassicalandNew_2003} The degree and handedness of the rotation are determined by the magnitude and the sign of $C_{15}$, respectively. Torsion only occurs when the nanowire bends along the $[\bar{1}10]$ direction independent of the shell geometry, confirming that it is the material symmetry which is responsible for this effect
\ifjournal
	(see Figs.~S2 and S3 in the supplementary material).
\else
	(see Figs.~\ref{fig:Appx_ZB_WZstrain} and \ref{fig:Appx_torsion} in the appendix).
\fi

We note that properties such as the bending or torsion of nanobeams are an area of very active research in the field of mechanical engineering, and are believed to require higher-order theories beyond classical linear elasticity, such as Cosserat, strain or stress gradient elasticity\mycite{.}{lakes_Int.J.Mech.Sci._2001, papanicolopulos_Int.J.SolidsStruct._2011, polizzotto_Eur.J.Mech.A-Solids_2015} The various models developed in this field aim to describe primarily the size dependence of deformations, and thus include length scale parameters to explicitly consider the actual size of the system under consideration\mycite{.}{askes_Int.J.SolidsStruct._2011,lurie_Int.J.Eng.Sci._2018} Such a size dependence may arise due to the microscopic heterogeneity of the material\mycite{.}{abali_Arch.Appl.Mech._2017} Still, in the extreme nanoscopic limit, an intrinsic size dependence is also expected for homogeneous materials (such as for single-crystalline GaAs nanowires), since the magnitude of the strain gradients linearly increases with decreasing system size\mycite{.}{li_J.Appl.Phys._2009,xu_Int.J.Appl.Mech._2013} However, there are two major issues with the existing higher-order elasticity theories: they offer conflicting predictions with no consensus in the field for the underlying reasons\mycite{,}{kaiser_Meccanica_2021} and for any practical application, they rely on a large number of additional and in general unknown parameters\mycite{.}{askes_Int.J.SolidsStruct._2011,lazar_Contin.Mech.Thermodyn._2022} Our model may be conceptually less advanced, but in the end more practical since all parameters involved are known with a fair degree of accuracy.

\ifjournal\else
	\section{Axial elastic strain and curvature}\label{sec:e_zz}
\fi
The full strain distribution in the bent nanowires as depicted in Fig.~\ref{fig:ZB_WZstrain} can only be obtained by numerical means. It is possible, however, to derive an analytical expression for the strain component $\elstrain_{zz}$ for a given shell thickness $\delta$ and core diameter $d$.
In \refcite{hadjimichael_Int.J.Numer.MethodsEng._2024}, we have taken the same approach to minimize the strain energy density in the nanowire, but modified the prestrain condition to ensure that the resulting expression coincides with the one obtained with the classical bending theory developed by Timoshenko\mycite{.}{timoshenko_J.Opt.Soc.Am._1925} This logical consistency is achieved only if we consider that the entire predeformation occurs in the shell, i.\,e., when $\M$ is the identity tensor in the core, and a diagonal matrix with elements corresponding to the full lattice mismatch in the shell\mycite{.}{hadjimichael_Int.J.Numer.MethodsEng._2024} In Fig.~\ref{fig:ezz_vs_y}, we compare the analytically derived variation of $\elstrain_{zz}$ with the results obtained from the numerical simulations. Figures \ref{fig:ezz_vs_y}(a) and \ref{fig:ezz_vs_y}(b) depict the variation of $\elstrain_{zz}$ along the center horizontal and vertical line across the nanowire cross sections shown in Figs.~\ref{fig:ZB_WZstrain}(a) and \ref{fig:ZB_WZstrain}(c), respectively. In both cases, the analytically derived $\elstrain_{zz}$ strain components are in excellent agreement with the numerical simulations.
\ifjournal
	\begin{figure}[t!]
		\centering
		\includegraphics[width=\columnwidth]{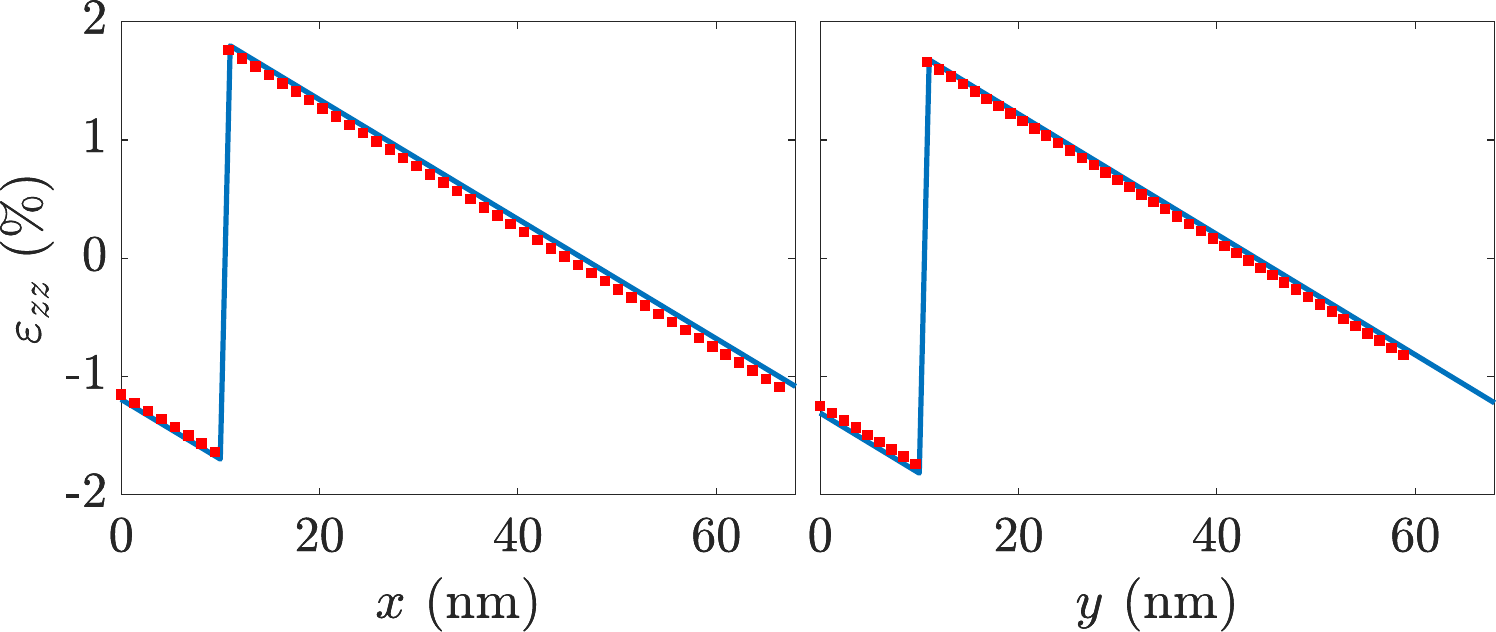}
\else
	\begin{figure}[bh!]
		\centering
		\includegraphics[width=0.75\columnwidth]{figures/csZBeZZ}
\fi
	\caption{Variation of $\elstrain_{zz}$ along the $x$ and $y$ directions of the ZB-[111] GaAs/$\text{In}_{0.5}\text{Al}_{0.5}\text{As}$ nanowire with $d = 100/\sqrt{3}$\,nm and $\delta = 10$\,nm as depicted in (left) Fig.~\ref{fig:ZB_WZstrain}(a) and (right) Fig.~\ref{fig:ZB_WZstrain}(c). The lines represent the analytical expression for $\elstrain_{zz}$\mycite{,}{hadjimichael_Int.J.Numer.MethodsEng._2024}, whereas the solid circles show the results of the finite-element simulation. Similar results are obtained for the WZ-[0001] case.}
	\label{fig:ezz_vs_y}
\end{figure}
\ifjournal
	\begin{figure}[b!]
		\centering
		\includegraphics[width=0.8\columnwidth]{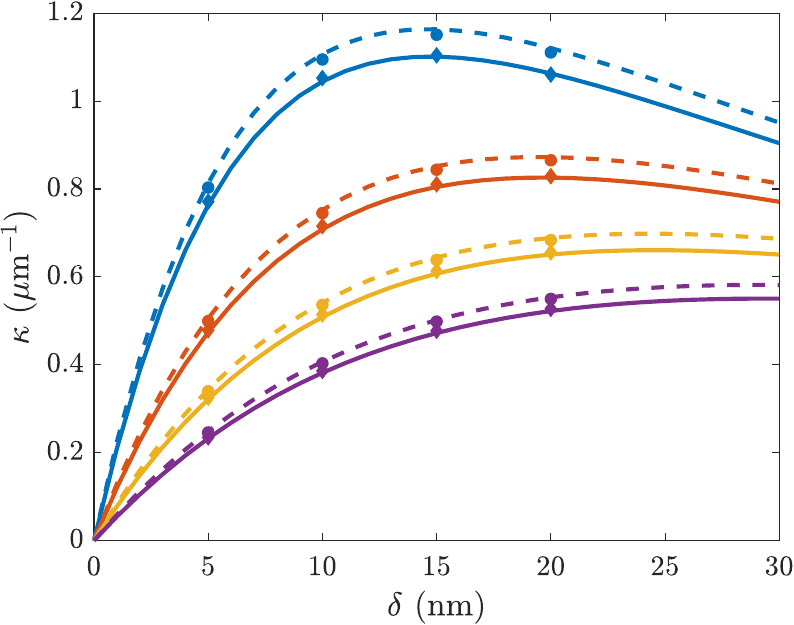}
\else
	\begin{figure}[bh!]
		\centering
		\includegraphics[width=0.6\columnwidth]{cs2curvature}
\fi
	\caption{Curvature of a bent GaAs/$\text{In}_{0.5}\text{Al}_{0.5}\text{As}$ nanowire with varying shell thickness $\delta$ for core diameters $d = \frac{1}{\sqrt{3}}\left\{60, 80, 100, 120\right\}$\,nm (blue, red, gold, and magenta curves, respectively). The solid and dashed lines are obtained with the analytical curvature formula for a ZB-[111] and WZ-[0001] nanowire, respectively. The solid circles and diamonds indicate the repective values obtained by the finite-element simulations. In both cases, the cross sections of Figs.~\ref{fig:ZB_WZstrain}(c) and \ref{fig:ZB_WZstrain}(d) have been used. Similar results are observed for the cross sections of the upper panel of Fig.~\ref{fig:ZB_WZstrain}.}
	\label{fig:nanowire_curvature}
\end{figure}
Based on the analytical expression for $\elstrain_{zz}$, we have derived a corresponding one for the curvature of the nanowire\mycite{.}{hadjimichael_Int.J.Numer.MethodsEng._2024} Figure~\ref{fig:nanowire_curvature} compares the curvature thus obtained as a function of $\delta$ and for several different values of $d$ with the one predicted by the numerical solution of \eqref{eq:divergence_energy_equilibrium} for both the ZB and the WZ structures. Evidently, the agreement between the analytical model based on $\elstrain_{zz}$ alone and the numerical model explicitly involving the full strain tensor is excellent, demonstrating that the bending of the nanowires is essentially exclusively determined by $\elstrain_{zz}$. The slightly larger values of $\elstrain_{zz}$ for the WZ-[0001] nanowires [cf.\ Figs.~\ref{fig:ZB_WZstrain}(a) and \ref{fig:ZB_WZstrain}(b), as well as Figs.~\ref{fig:ZB_WZstrain}(c) and \ref{fig:ZB_WZstrain}(d)] thus manifests itself in a systematically larger curvature.

For each core diameter, the curvature peaks at a certain shell thickness, the value of which is monotonically decreasing with decreasing core diameter. This behavior is easily understood: the smaller the core diameter, the easier it is for a mismatched shell of given thickness to bend the nanowire. With increasing shell thickness, the curvature initially increases until the shell becomes so thick that it switches roles with the core. From this point on, which is inversely proportional to the core diameter, it is the shell which is decreasingly bend by a core of constant diameter.

\ifjournal\else
	\section{Outlook: consequences for piezoelectricity}\label{sec:piezoelectricity}
\fi
Since GaAs is a piezoelectric material, it is important to consider the consequences of the distinct strain fields of the two polytypes for piezoelectric polarization and piezoelectric fields. The piezoelectric polarization $\bm{P} = (P_x, P_y, P_z)^T$ for the WZ polytype in [0001] orientation has the form\mycite{.}{schulz_Phys.Rev.B_2011}
\begin{align*}
    \bm{P}^{0001} = \begin{pmatrix}[1.2]
                    2e^{\text{\tiny WZ}}_{15}\elstrain_{xz} \\
                    2e^{\text{\tiny WZ}}_{15}\elstrain_{yz} \\
                    2e^{\text{\tiny WZ}}_{31}(\elstrain_{xx}+\elstrain_{yy})+e^{\text{\tiny WZ}}_{33}\elstrain_{zz}
                    \end{pmatrix},
\end{align*}
where the $e^{\text{\tiny WZ}}_{ij}$ are the piezoelectric constants for the WZ structure
\ifjournal
	(see Tab.~S1 in the supplementary material).
\else
	(see Tab.~\ref{tab:Appx_parameters} in the appendix).
\fi
Since $\elstrain_{xz} = \elstrain_{yz} = 0$, it follows that $P_x = P_y = 0$, i.\,e., there is neither an in-plane piezoelectric polarization, nor a transverse piezoelectric field. Only the polarization component $P_z$ does not vanish for the WZ-[0001] nanowires. However, since $P_z$ does not change along $z$ (except in direct vicinity of the bottom interface with the substrate and the free top surface of the nanowire), this polarization does not induce any longitudinal piezoelectric field.

For the ZB polytype in [111] orientation, the polarization vector in the rotated basis reads\mycite{}{schulz_Phys.Rev.B_2011}
\ifjournal
	{\small
\fi
\begin{align*}
    \bm{P}^{111} = \begin{pmatrix}[1.2]
                    2e^{\text{\tiny ZB}}_{15}\elstrain_{xz} \\
                    2e^{\text{\tiny ZB}}_{15}\elstrain_{yz} \\
                    2e^{\text{\tiny ZB}}_{31}(\elstrain_{xx}+\elstrain_{yy})+e^{\text{\tiny ZB}}_{33}\elstrain_{zz}
                    \end{pmatrix}
                    +
                    \begin{pmatrix}[1.2]
                    e^{\text{\tiny ZB}}_{11}(\elstrain_{xx}-\elstrain_{yy}) \\
                    2e^{\text{\tiny ZB}}_{12}\elstrain_{xy} \\
                    0
                    \end{pmatrix},
\end{align*}
\ifjournal
	}
\fi
where the $e^{\text{\tiny ZB}}_{ij}$ are the piezoelectric constants for the ZB-[111] case (for their relation to the single piezoelectric constant $e_{14}$ of the ZB crystal in its natural $\langle 001 \rangle$ basis, see \refcite{schulz_Phys.Rev.B_2011} and
\ifjournal
	Tab.~S1 in the supplementary material).
\else
	Tab.~\ref{tab:Appx_parameters} in the appendix).
\fi
In contrast to the case of WZ-[0001] nanowires discussed above, the polarization components $P_x$ and $P_y$ are not zero, and neither are $\partial P_x/\partial x$ and $\partial P_y/\partial y$
\ifjournal
	(see Fig.~S4 in the supporting material
\else
	(see Fig.~\ref{fig:Appx_piezoelecticity} in the appendix
\fi
for the corresponding polarization maps). Hence, in bent ZB-[111] nanowires, we encounter transverse electric fields of substantial magnitude.

\ifjournal\else
	\section{Final remarks}\label{sec:conclusion}
\fi
In summary and conclusion, our analysis of the strain distribution in GaAs nanowires bend by a one-sided stressor shell has revealed several important differences between the ZB and WZ polytypes. First of all, for the ZB-[111] case, all shear strain components are large, whereas  the out-of-plane ones are zero for otherwise equivalent WZ-[0001] structures. The origin of this difference lies in the fact that the stiffness matrix for the ZB structure, rotated such as to align with the nanowire coordinates, contains a non-zero symmetry-breaking (chiral) stiffness constant $C_{15}$. The resulting large out-of-plane shear strain components leads to a torsion for a certain bending direction, and thus makes these nanowires chiral.

Moreover, these shear strain components for the ZB case directly translate into large piezoelectric polarization components $P_x$ and $P_y$, which, due to their variation along $x$ and $y$, will cause non-negligible transverse electrostatic fields. These fields may be strong enough to field-ionize excitons and spatially separate charge carriers, and thus to profoundly modify the charge carrier dynamics in these nanowires. To understand this dynamics, we are required to solve the drift-diffusion equations for electrons and holes in the potential landscape created by strain as well as by the resulting piezo- und flexoelectric potentials. As a first step, we are currently extending our model to include full electromechanical coupling for being able to compute the total electrostatic potential in nanowires bent by one-sided stressor shells.

\ifwias\else
	\ifarxiv
		\section*{Acknowledgments}
	\fi
	This work is supported by the Leibniz competition 2020 (NUMSEMIC, J89/2019) and the Deutsche
	Forschungsgemeinschaft (DFG) (project number: 434114264).
	We thank Oliver Marquardt and Taseer Anjum (Universität Siegen) for valuable discussions, and Vladimir
	Kaganer for a critical reading of the manuscript.
\fi

\ifjournal\else
	\pagebreak
	\appendix
	\setcounter{figure}{0}
	\setcounter{table}{0} 
	\renewcommand\thefigure{S\arabic{figure}}
	\renewcommand\thetable{S\arabic{table}} 
	\section*{Appendix}
	This appendix provides additional details on the bending of nanowires with one-sided stressor shells. In particular, we examine the strain and the torsion of bent ZB and WZ nanowires that are assumed to have $\{11\bar{2}\}$ and $\{1\bar{1}00\}$ facets, respectively, hence having the shell applied in the crystallographically opposite directions compared to the nanowires in the main text. Furthermore, we present the polarization potential for the nanowires examined in the main text. Finally, we provide the material parameters used throughout our study.
	
	\section{Nanowire bending}
	Figure~\ref{fig:Appx_nanowire3D} provides a three-dimensional view of a 2~\textmu m long GaAs/$\text{In}_{0.5}\text{Al}_{0.5}\text{As}$ core/shell nanowire with a core diameter $d = 100/\sqrt{3}$\,nm and a shell thickness $\delta =10$\,nm. The zoomed-in inlay shows the deformed position of a cross section taken at at $1.6$~\textmu m from the nanowire base, and the crystallographic directions for the ZB-[111] case. The curvature $\kappa$ of the bent nanowire is $0.51$\,\textmu m$^{-1}$.
	\begin{figure}[b!]
		\centering
		\includegraphics[width=0.7\textwidth]{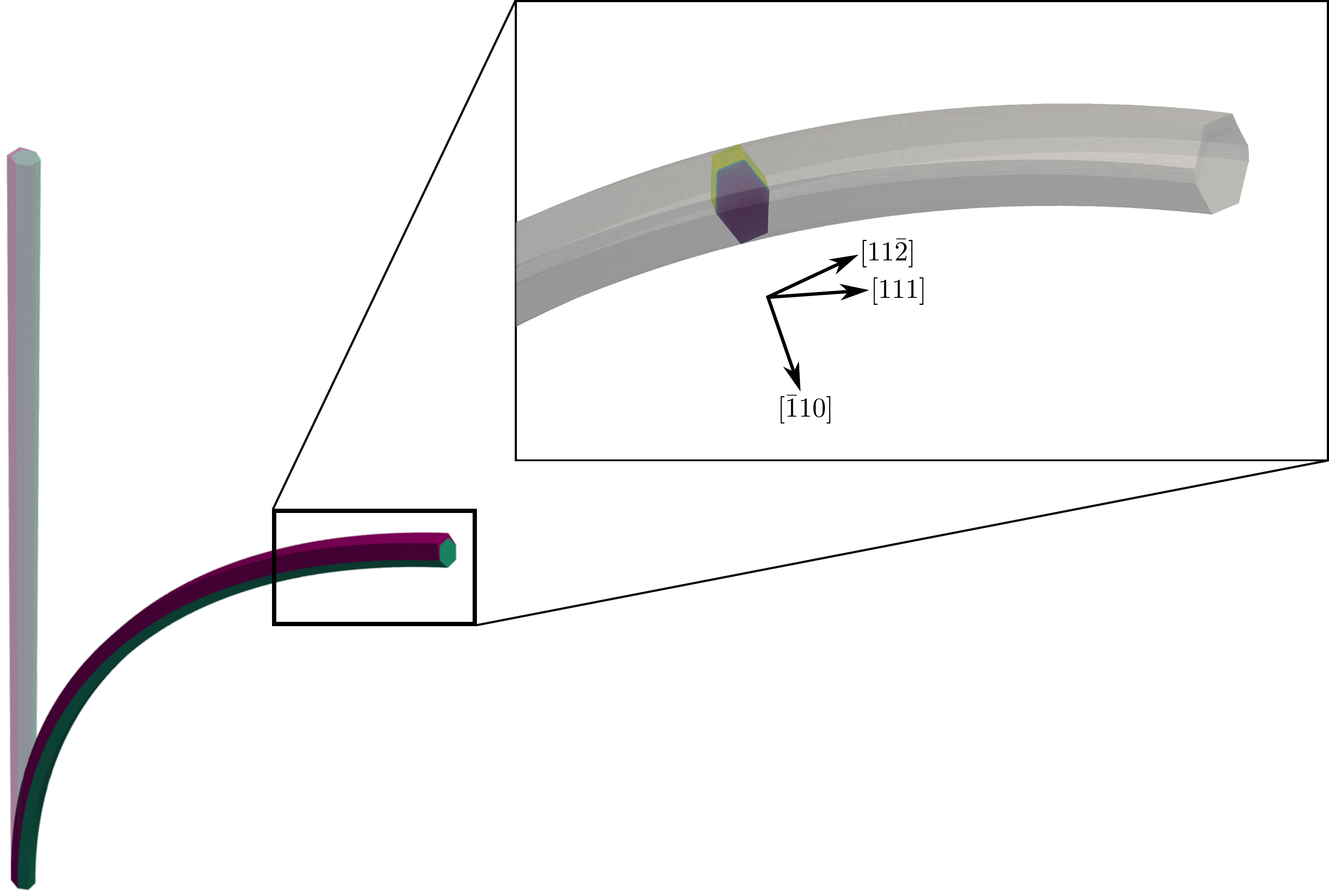}
		\caption{Unbent and bent configuration of a $2000$\, nm long GaAs/$\text{In}_{0.5}\text{Al}_{0.5}\text{As}$ core/shell nanowire with $100/\sqrt{3}$\,nm core diameter and $10$\,nm shell thickness. Magnified view of the nanowire with positions of deformed cross sections at height $z = 1600$\,nm (on the unbent configuration) of a ZB-111 nanowire. At the cross section, the spatial coordinate axes $(x, y, z)$ are aligned with the $([\bar{1}10], [11\bar{2}], [111])$ crystallographic directions.}
		\label{fig:Appx_nanowire3D}
	\end{figure}
	
	\section{Strain distribution and torsion}
	\begin{figure}[bh!]
		\centering
		\includegraphics[width=0.49\textwidth]{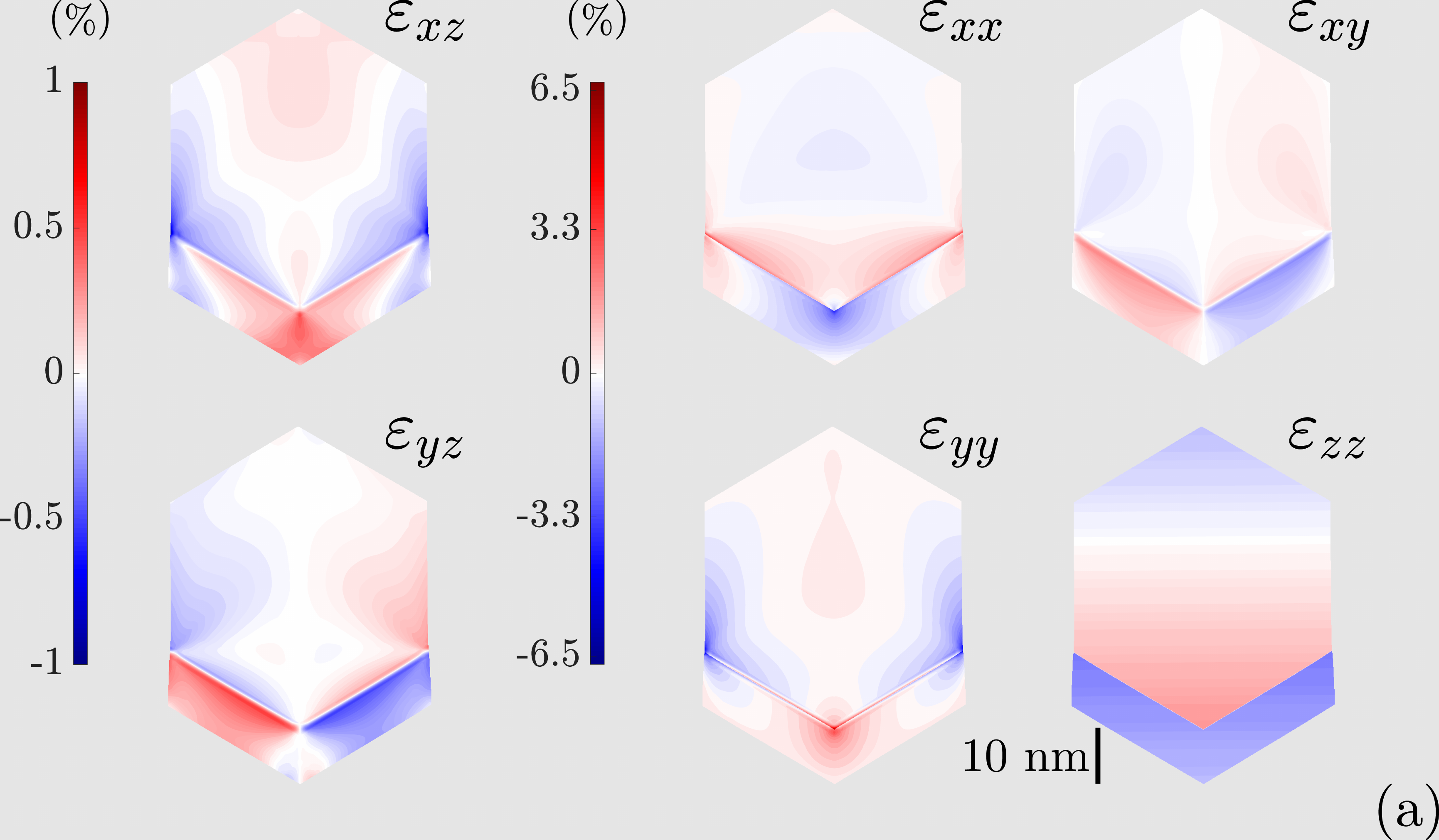}\hfill
		\includegraphics[width=0.49\textwidth]{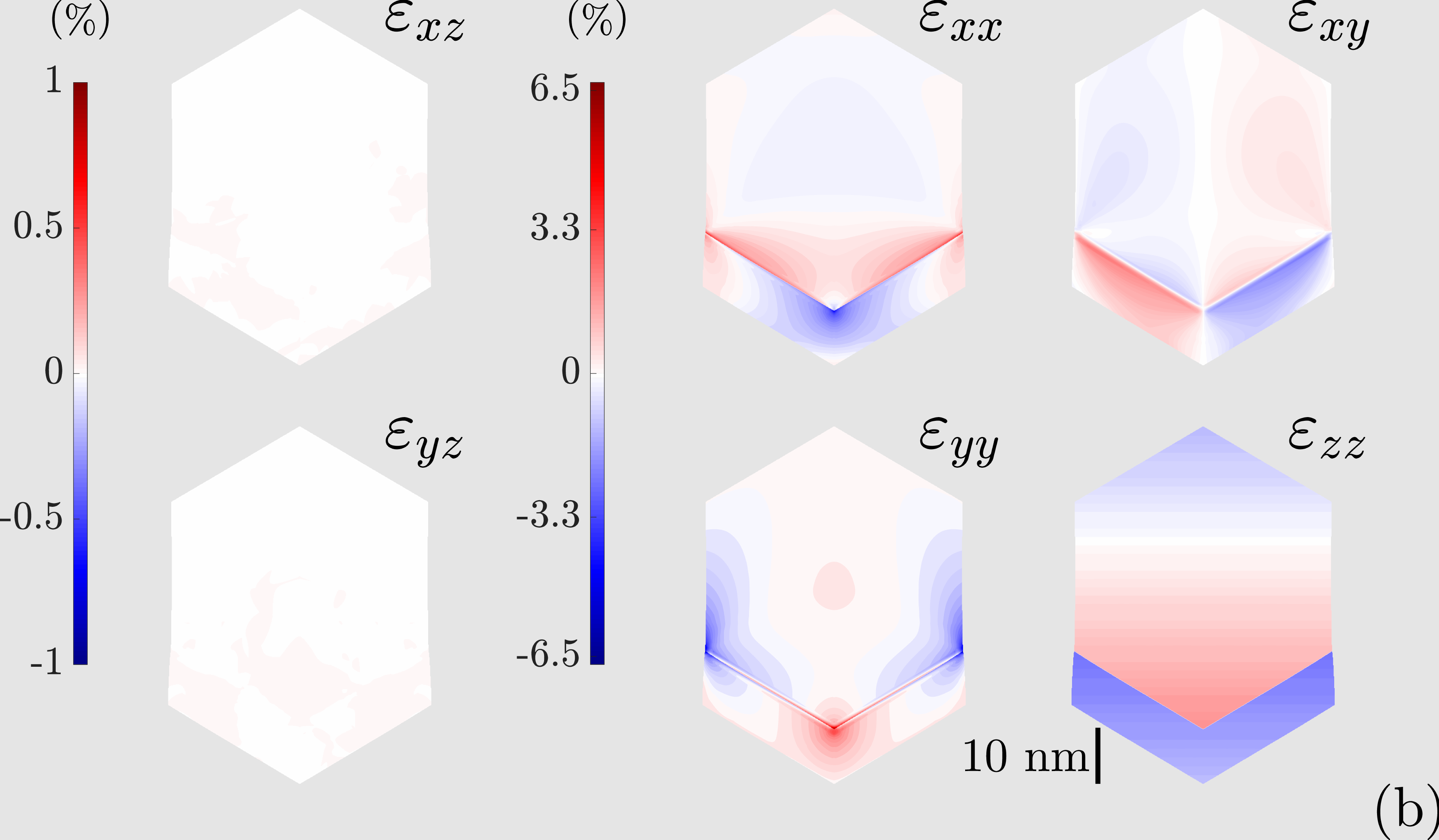}\\[10pt]
		\includegraphics[width=0.49\textwidth]{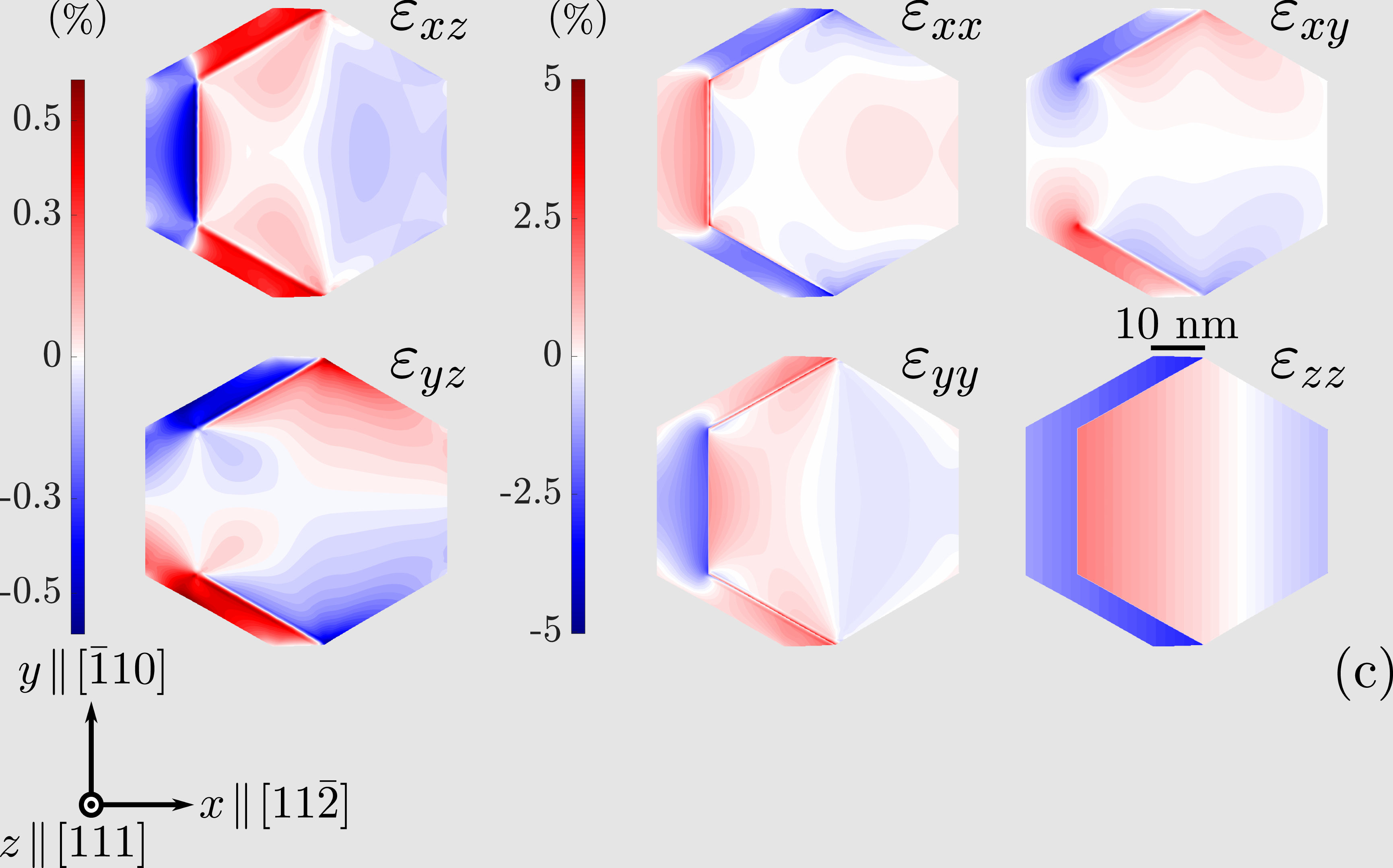}\hfill
		\includegraphics[width=0.49\textwidth]{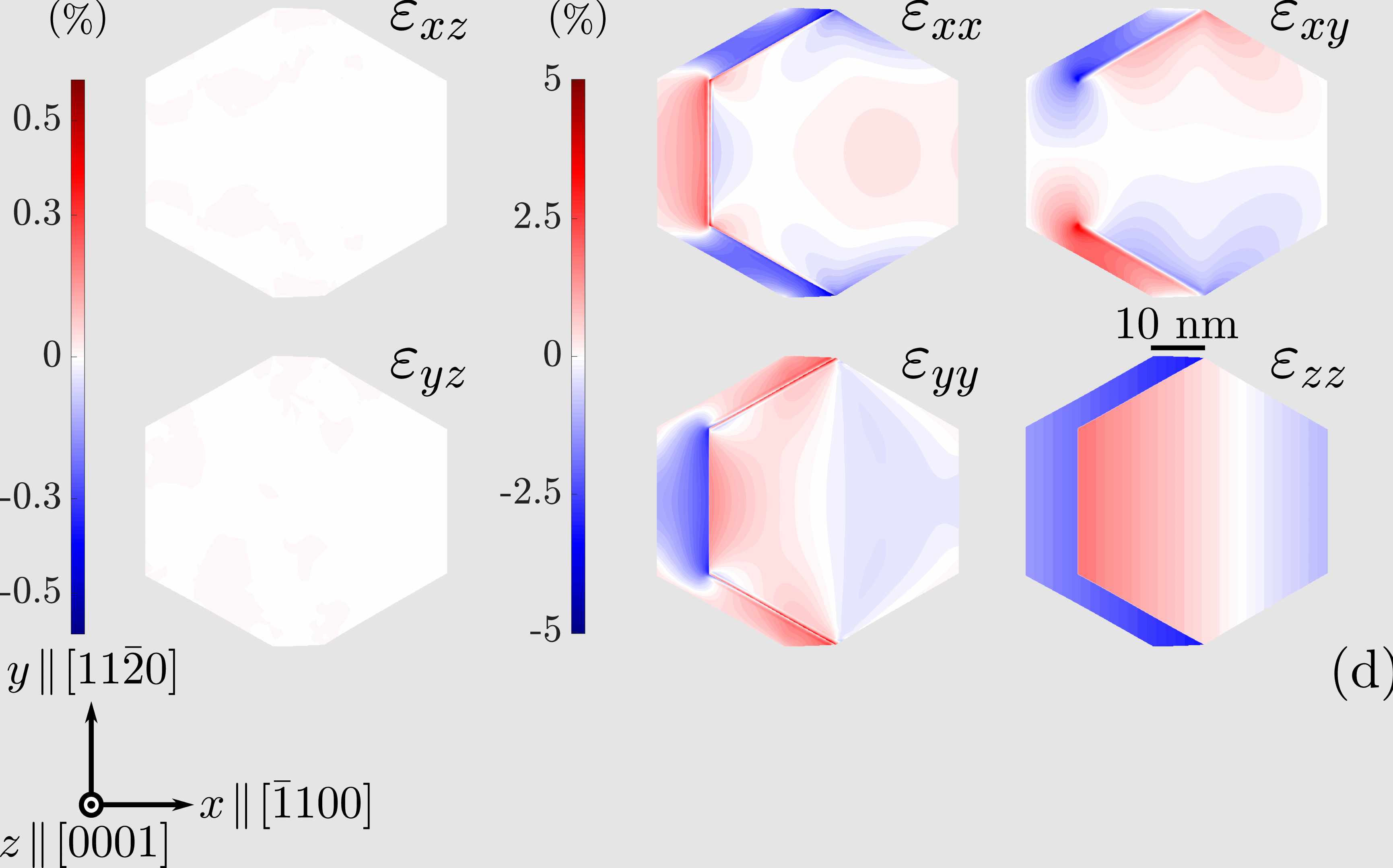}
		\caption{Elastic strain components computed on the central cross section of a 2~\textmu m long GaAs/$\text{In}_{0.5}\text{Al}_{0.5}\text{As}$ core/shell nanowire, using two different cross-sectional geometries as in Fig.~\ref{fig:ZB_WZstrain}. Here, the GaAs core is crystallographically rotated by 90°, and the shell is thus applied in opposite directions compared to Fig.~\ref{fig:ZB_WZstrain}. In the top panel, the shell is applied on a vertex of the hexagonal core (a) along the $[\bar{1}10]$ direction for ZB-[111] nanowires, and (b) along the $[11\bar{2}0]$ direction for WZ-[0001] nanowires. In the bottom panel, the shell is applied on the edge of the hexagonal core, i.\,e., in (c), along the $[11\bar{2}]$ direction for ZB-[111] nanowires, and in (d) along the $[\bar{1}100]$ direction for WZ-[0001] nanowires. In each case, we separate the $\elstrain_{xz}$ and $\elstrain_{yz}$ shear strain components since their values are smaller than the normal components. In the ZB-111 case [see (a) and (c)], the $\elstrain_{xz}$ and $\elstrain_{yz}$ components are approximately $6.5$ and $10$ times smaller than those of the normal strain, but in the WZ-[0001] case [see (b) and (d)], they are zero.}
		\label{fig:Appx_ZB_WZstrain}
	\end{figure}
	Figure~\ref{fig:Appx_ZB_WZstrain} shows the six elastic strain components in the center cross-section (taken at 1~\textmu m) of a 2~\textmu m long GaAs/$\text{In}_{0.5}\text{Al}_{0.5}\text{As}$ core/shell nanowire with a core diameter $d = 100/\sqrt{3}$\,nm, and a shell thickness $\delta =10$\,nm. We consider two different polytypes [ZB in (a) and (c), WZ in (b) and (d)] and two different core/shell geometries (the two top and bottom rows). In contrast to Fig.~\ref{fig:ZB_WZstrain}, the ZB (WZ) GaAs core is assumed to have ${11\bar{2}}$ (${1\bar{1}00}$) facets allowing the shell to be applied in opposite directions as compared to Fig.~\ref{fig:ZB_WZstrain}. The strain distribution shows a similar behavior to the one observed in the main text. In all four cases, the normal strain components $\elstrain_{xx}$, $\elstrain_{yy}$, and $\elstrain_{zz}$ are large and very similar in magnitude for both polytypes and crystallographic bending directions. The same applies to the in-plane shear strain component $\elstrain_{xy}$.
	\begin{figure}[b!]
		\centering
		\includegraphics[width=0.3\textwidth]{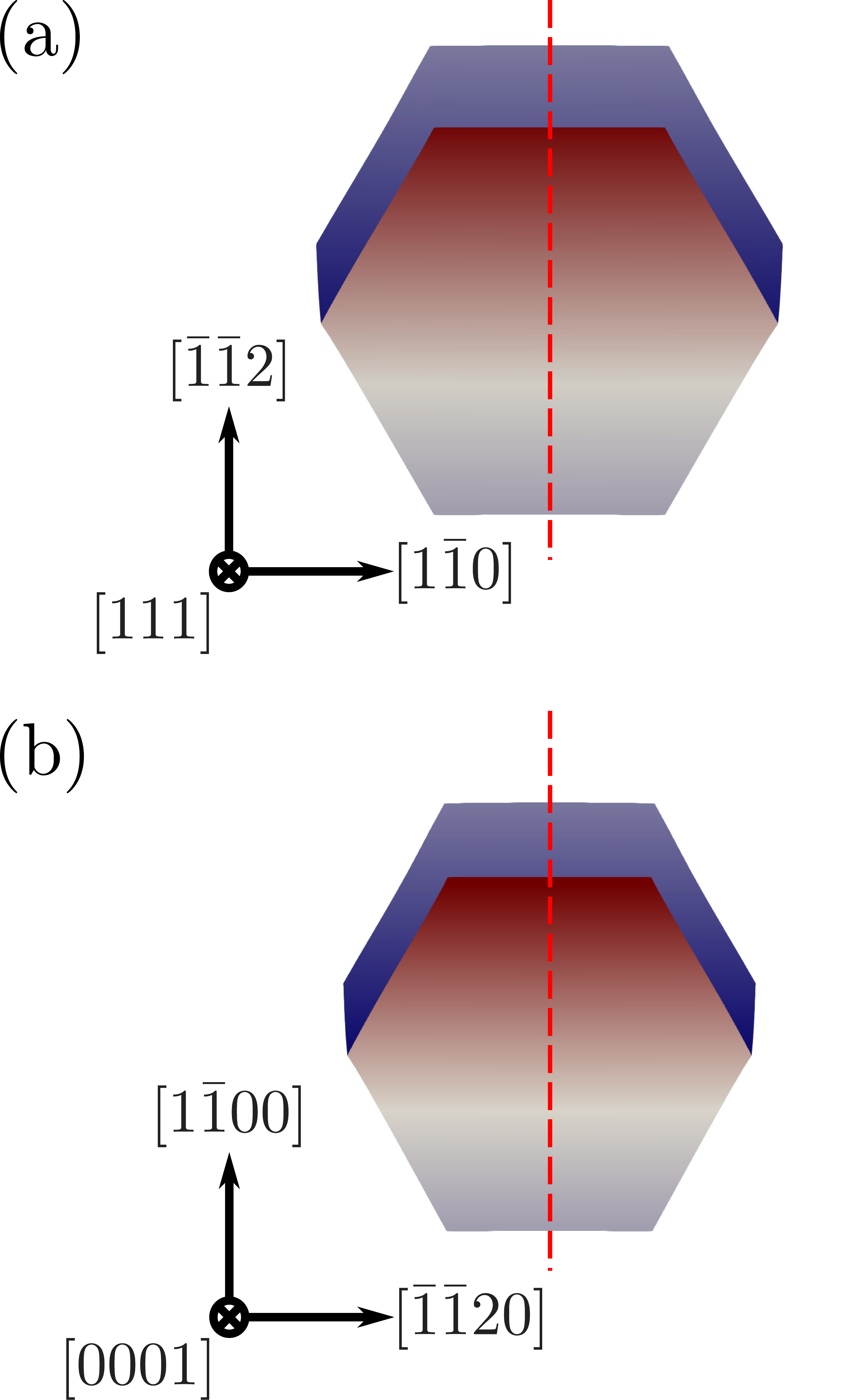}\qquad
		\includegraphics[width=0.3\textwidth]{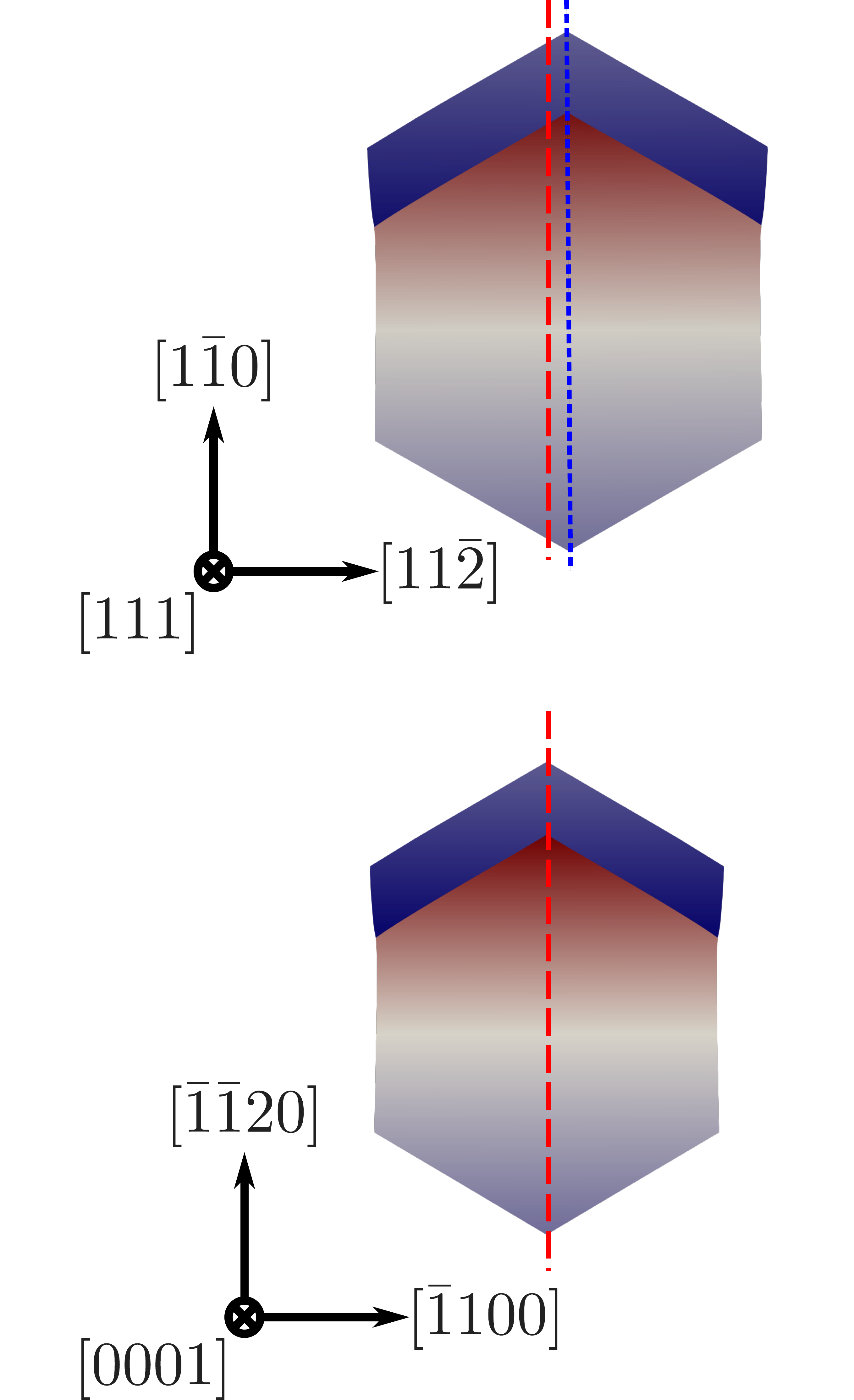}
		\caption{Bent GaAs/InAs nanowires viewed along the normal of their top facet from the same viewpoint (inside the nanowire). Due to the different curvature and thus distance from the viewpoint, the nanowires seem to be of different size, but they all have an identical core diameter of $100/\sqrt{3}$\,nm. The color gradient depicts the variation of $\elstrain_{zz}$ analogously to Fig.~\ref{fig:Appx_ZB_WZstrain}. In (a), we show ZB-[111] nanowires with the stressor shell deposited along the (left) $[11\bar{2}]$ and (right) $[\bar{1}10]$ directions. In (b), the corresponding WZ-[0001] nanowires are displayed with the stressor shell deposited along the (left) $[\bar{1}100]$ and (right) $[11\bar{2}0]$ directions. The red dashed line represents the bending plane for each of the four different cases. In three of these cases, the nanowire stays exactly in the bending plane. However, for ZB-[111] nanowires with the shell applied in the $[\bar{1}10]$ direction, the nanowires is laterally displaced and exhibits a torsion visualized by the blue dotted line.}
		\label{fig:Appx_torsion}
	\end{figure}
	Figure~\ref{fig:Appx_torsion} shows the 90° rotated nanowires along the normal of their top facet for our four different bent nanowire configurations. To maximize the displacements, we here consider GaAs nanowires with a pure InAs stressor shell, representing the maximum lattice mismatch available in this materials system. For the WZ-[0001] nanowires shown in Fig.~\ref{fig:Appx_torsion}(b), the nanowire bends strictly in the bending plane indicated by the red dashed lines. The same is true for the ZB-[111] nanowire shown in the left panel of Fig.~\ref{fig:Appx_torsion}(a) with the stressor shell applied along the $[11\bar{2}]$ direction. However, for the ZB-[111] nanowire shown in the right panel of Fig.~\ref{fig:Appx_torsion}(a) with the stressor shell applied along the $[\bar{1}10]$ direction, we observe a displacement of the nanowire out of the bending plane, and an anticlock-wise rotation about the nanowire's [111] axis. Note that the torsion observed in this nanowire configuration is significantly smaller than the one showed in Fig.~\ref{fig:torsion}(a), and the magnitude of the torsion primarily depends on whether the shell is applied on a vertex or the edge of the hexagonal core [see right column of Fig.~\ref{fig:torsion}(a) and Fig.~\ref{fig:Appx_torsion}(a)].
	
	\section{Piezoelectricity}
	Figure~\ref{fig:Appx_piezoelecticity} shows the polarization components $P_x$, $P_y$, and $P_z$ for the nanowire configurations of Fig.~\ref{fig:ZB_WZstrain} of the main manuscript. For the ZB-[111] case, the polarization components $P_x$ and $P_y$ are not zero, and neither are $\partial P_x/\partial x$ and $\partial P_y/\partial y$. Hence, the bent nanowire encounters transverse electric fields of substantial magnitude. In contrast, the $P_x$ and $P_y$ are zero in the WZ-[0001] case; thus there is neither an in-plane piezoelectric polarization, nor a transverse piezoelectric field. Only the polarization component $P_z$ does not vanish for the WZ-[0001] nanowires. However, since $P_z$ does not change along $z$ (except in direct vicinity of the bottom interface with the substrate and the free top surface of the nanowire), this polarization does not induce any longitudinal piezoelectric field.
	\begin{figure}[th!]
		\centering
		\includegraphics[width=0.5\textwidth]{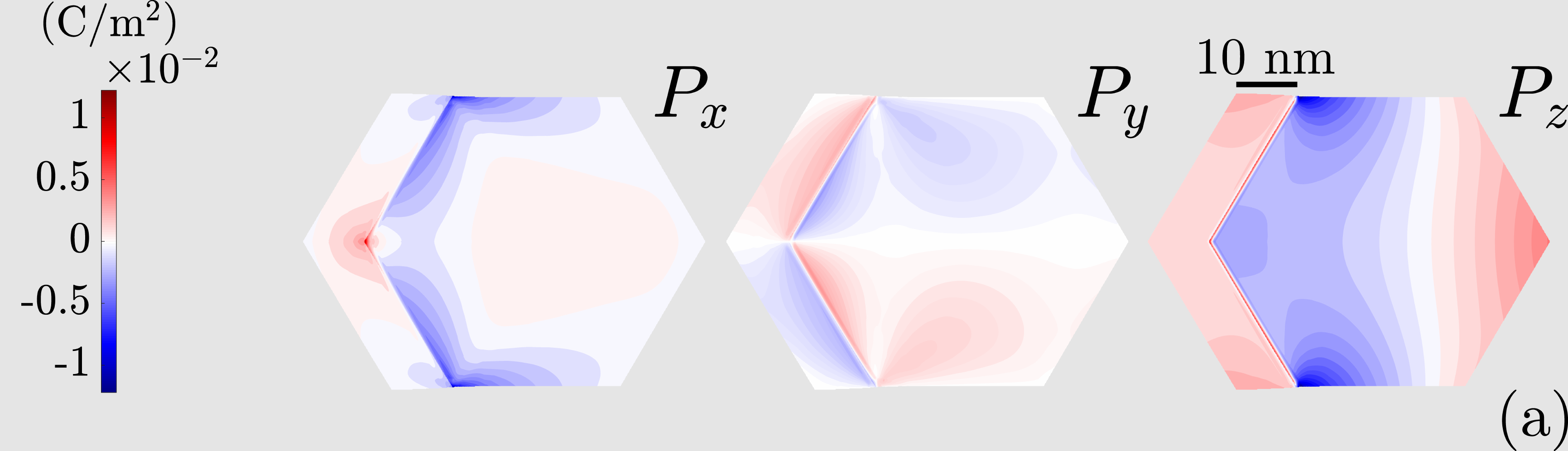}\hfill
		\includegraphics[width=0.5\textwidth]{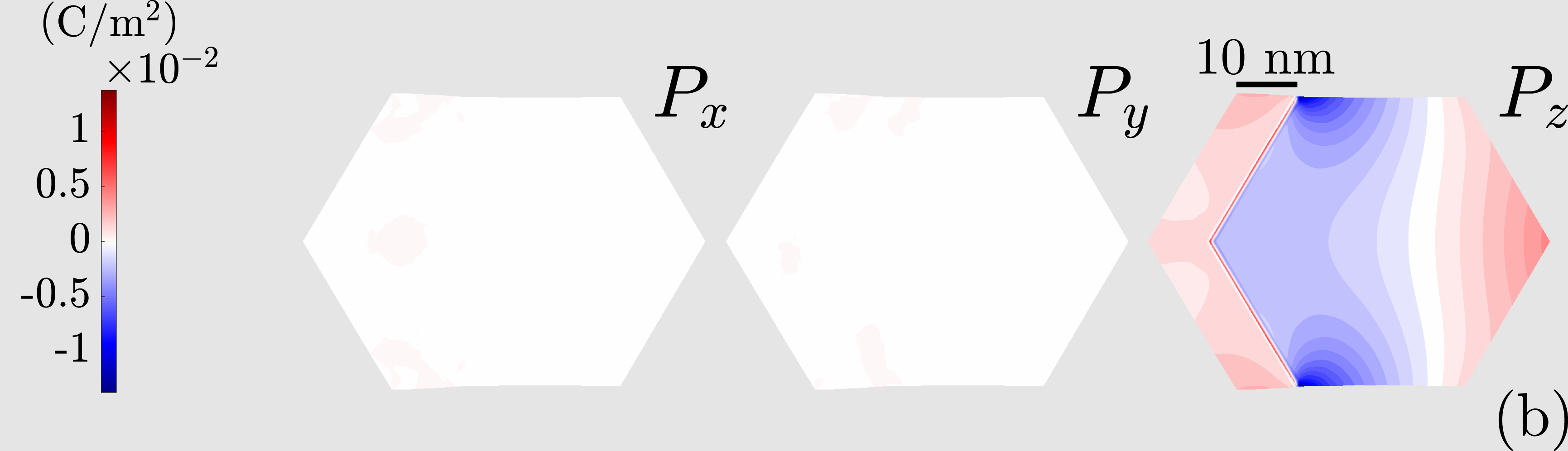} \\[10pt]
		\includegraphics[width=0.49\textwidth]{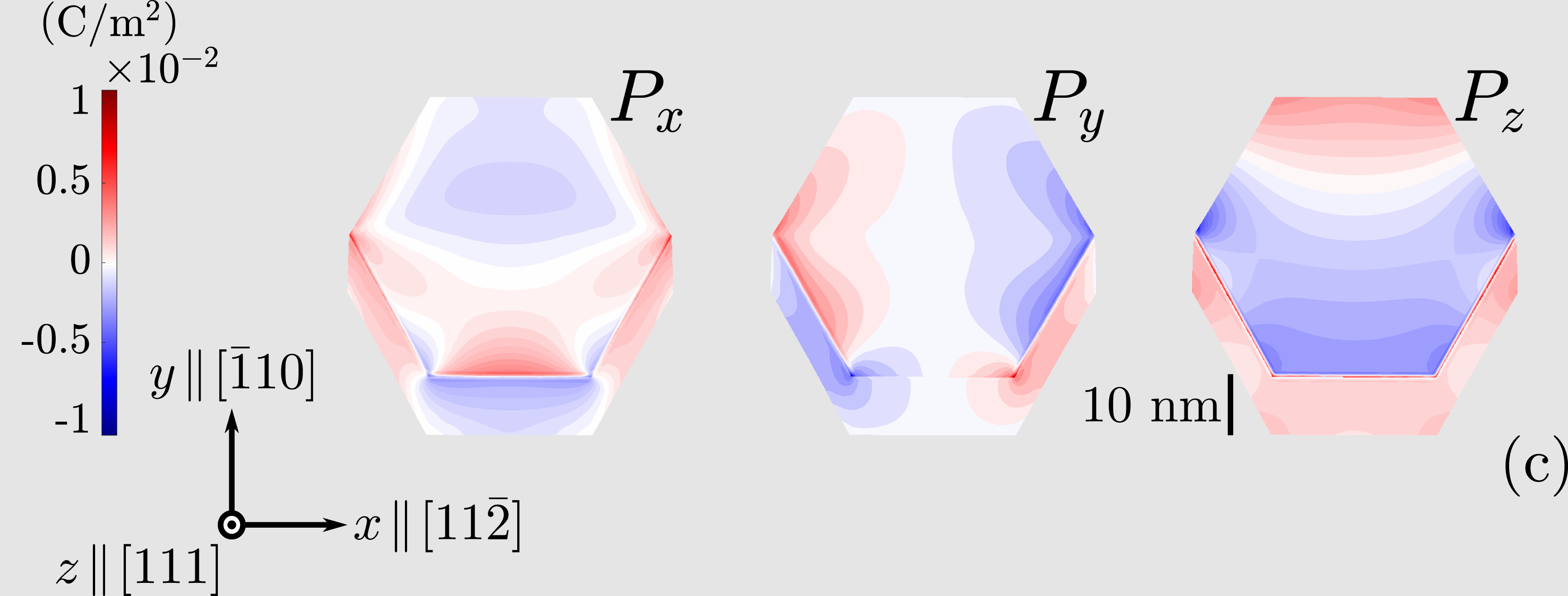}\hfill
		\includegraphics[width=0.49\textwidth]{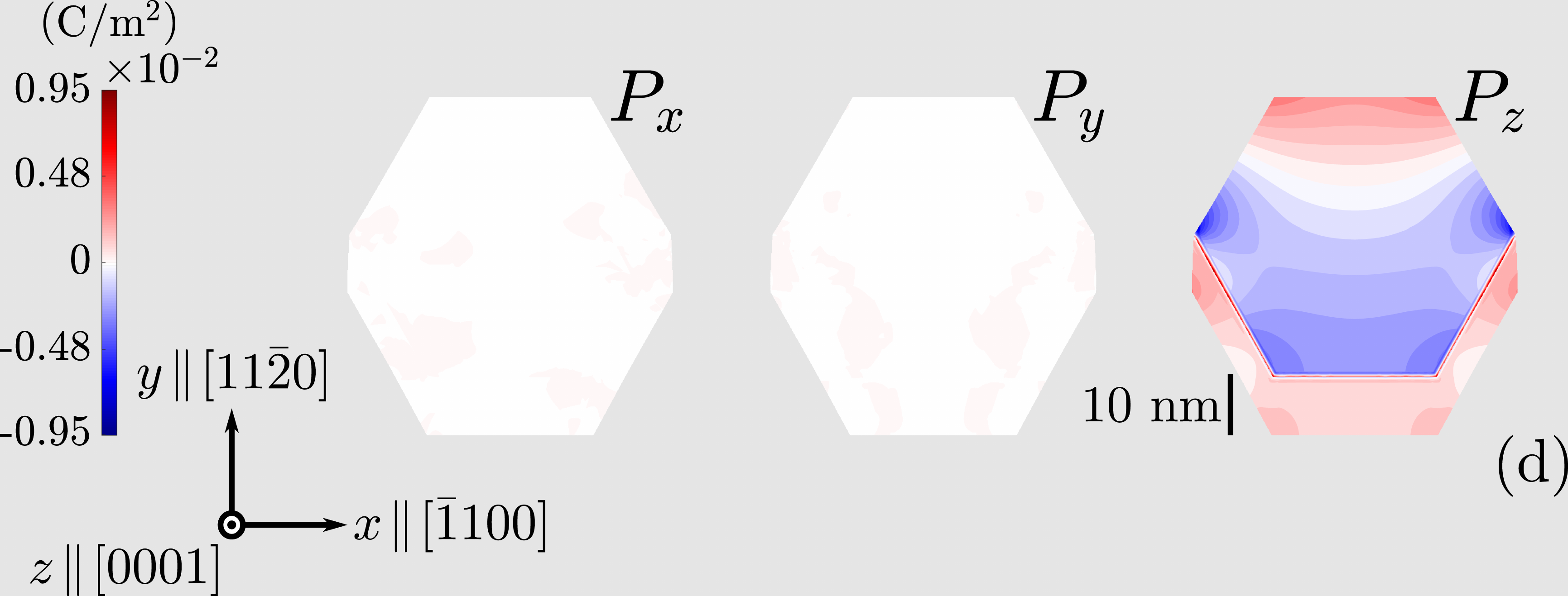}
		\caption{(a) Polarization field for a ZB-[111] nanowire bending in the $[11\bar{2}]$ direction [see
			Fig.~\ref{fig:ZB_WZstrain}(a)].
			(b) Polarization field for a WZ-[0001] nanowire bending in the $[\bar{1}100]$ direction [see Fig.~\ref{fig:ZB_WZstrain}(b)].
			(c) Polarization field for a ZB-[111] nanowire bending in the $[\bar{1}10]$ direction [see Fig.~\ref{fig:ZB_WZstrain}(c)].
			(d) Polarization field for a WZ-[0001] nanowire bending in the $[11\bar{2}0]$ direction [see
			Fig.~\ref{fig:ZB_WZstrain}(d)].}
		\label{fig:Appx_piezoelecticity}
	\end{figure}
	
	\section{Material parameters}
	Table~\ref{tab:Appx_parameters} shows the material parameters used throughout this study. Since there is no complete set of parameters available for the WZ materials, particularly for InAs and AlAs, we use the quasi-cubic approximation to obtain these parameters from those of the corresponding ZB materials.
	\begin{table*}
	  \begin{threeparttable}
		\centering
		\ifjournal
			\renewcommand{\arraystretch}{1}
		\else
			\renewcommand{\arraystretch}{1.5}
		\fi	
		\setlength{\tabcolsep}{5pt}
		\begin{tabular}{ccccccc}
		\hline
		\hline
		\multirow{2}{*}{\parbox{3cm}{\centering Parameter (unit)}} & \multicolumn{2}{c}{GaAs} &
			\multicolumn{2}{c}{InAs} & \multicolumn{2}{c}{AlAs} \\
		& ZB-[111] & WZ-[0001] & ZB-[111] & WZ-[0001] & ZB-[111] & WZ-[0001] \\
		\hline
		\hline
		$a, b$ (\si{\angstrom}) & $3.997$\tnote{a} & $3.994$\tnote{b} &
			$4.284$\tnote{a} & $4.311$\tnote{b} & $4.003$\tnote{a} & $4.002$\tnote{b} \\
		$c$ (\si{\angstrom}) & $6.5278$\tnote{a} & $6.586$\tnote{b} &
			$6.996$\tnote{a} & $7.093$\tnote{b} & $6.537$\tnote{a} & $6.582$ \\
		$C_{11}$ (GPa) & \multicolumn{2}{c}{$149.35$\tnote{a}} & \multicolumn{2}{c}{$103.87$\tnote{a}} &
			\multicolumn{2}{c}{$143.4$\tnote{a}} \\
		$C_{12}$ (GPa) & \multicolumn{2}{c}{$47.52$\tnote{a}} & \multicolumn{2}{c}{$38.40$\tnote{a}} &
			\multicolumn{2}{c}{$47.27$\tnote{a}} \\
		$C_{13}$ (GPa) & \multicolumn{2}{c}{$38.43$\tnote{a}} & \multicolumn{2}{c}{$31.54$\tnote{a}} &
			\multicolumn{2}{c}{$41.13$\tnote{a}} \\
		$C_{15}$ (GPa) & $12.85$\tnote{a} & $0$ & $9.70$\tnote{a} & $0$ & $8.67$\tnote{a} & $0$ \\
		$C_{33}$ (GPa) & \multicolumn{2}{c}{$158.43$\tnote{a}} & \multicolumn{2}{c}{$110.72$\tnote{a}} &
			\multicolumn{2}{c}{$149.53$\tnote{a}} \\
		$C_{44}$ (GPa) & \multicolumn{2}{c}{$41.83$\tnote{a}} & \multicolumn{2}{c}{$25.87$\tnote{a}} &
			\multicolumn{2}{c}{$41.93$\tnote{a}} \\
		$C_{66}$ (GPa) & \multicolumn{2}{c}{$50.92$\tnote{a}} & \multicolumn{2}{c}{$32.73$\tnote{a}} &
			\multicolumn{2}{c}{$48.07$\tnote{a}} \\	
		$e_{11}$ (C/m$^{2}$) & $ 0.194$\tnote{c} & $0$ & $ 0.094$\tnote{c} & $0$ & $ 0.039$\tnote{c} & $0$ \\
		$e_{12}$ (C/m$^{2}$) & $-0.194$\tnote{c} & $0$ & $-0.094$\tnote{c} & $0$ & $-0.039$\tnote{c} & $0$ \\
		$e_{15}$ (C/m$^{2}$) & \multicolumn{2}{c}{$0.137$\tnote{c}} & \multicolumn{2}{c}{$0.066$\tnote{c}} &
			\multicolumn{2}{c}{$0.028$\tnote{c}} \\
		$e_{31}$ (C/m$^{2}$) & \multicolumn{2}{c}{$0.137$\tnote{c}} & \multicolumn{2}{c}{$0.066$\tnote{c}} &
		\multicolumn{2}{c}{$0.028$\tnote{c}} \\
		$e_{33}$ (C/m$^{2}$) & \multicolumn{2}{c}{$-0.274$\tnote{c}} & \multicolumn{2}{c}{$-0.133$\tnote{c}} &
		\multicolumn{2}{c}{$-0.055$\tnote{c}} \\
		\hline
		\hline
		\end{tabular}
		\begin{tablenotes}
			\item [a] Ref.~\cite{vurgaftman_J.Appl.Phys._2001}.
			\item [b] Ref.~\cite{jain_APL.Mater._2013}.
			\item [c] Quasi-cubic approximation (see \refcite{bernardini_Phys.Rev.B_1997}) using the
				value of $e_{14}^\text{\tiny ZB}$ in the natural coordinate system ($\langle 001 \rangle$) of the
				ZB crystal, taken from \refcite{beya-wakata_Phys.Rev.B_2011}
				($e_{11} = -\sqrt{\frac{2}{3}}e_{14}^\text{\tiny ZB}$,
				$e_{12} = -e_{11}$,
				$e_{15} = -\frac{1}{\sqrt{3}}e_{14}^\text{\tiny ZB}$,
				$e_{31} = e_{15}$,
				$e_{33} = \frac{2}{\sqrt{3}}e_{14}^\text{\tiny ZB}$).
				The $e_{14}^\text{\tiny ZB}$ values for \{GaAs, InAs, AlAs\} are $\{-0.238, -0.115, -0.048\}$
				C/m$^{2}$, respectively.
	    \end{tablenotes}
		\end{threeparttable}
		\caption{Material parameters of GaAs, InAs, and AlAs for the ZB-[111] and WZ-[0001] crystal
			structures.
			The parameters for the $\text{In}_{x}\text{Al}_{1-x}\text{As}$ shell are determined by a convex
			combination between the values of InAs and AlAs.}
		\label{tab:Appx_parameters}
	\end{table*}
\fi

\sloppy
\ifnum0%
    \ifjournal 1 \fi \ifwias 1 \fi > 0
    \ifjournal
        \bibliography{shortjournals,bibliography} 
    \else
        \bibliographystyle{agufull}
        \bibliography{longjournals,bibliography} 
    \fi
\else
    \printbibliography
\fi

\end{document}